\documentclass[
a4paper,
amsmath,
amssymb,
aps,
prd,
nofootinbib,
preprintnumbers,
longbibliography
]
{revtex4-2}
\usepackage[table]{xcolor} 
\usepackage{graphicx}
\usepackage{dcolumn}
\usepackage{bm}
\usepackage{tcolorbox}
\begin{document} 

\preprint{MPP-2025-234 01.06.2026}

\title{How to identify the dead cone in the top-quark jet}

\author{Stefan Kluth}
\author{Wolfgang Ochs}
\affiliation{Max-Planck-Institut f\"ur Physik, Boltzmannstr.\ 8, 85748 Garching, Germany}
\author{Redamy Perez-Ramos}
\affiliation{DRII-IPSA, Bis, 63 Boulevard de Brandebourg, 94200 Ivry-sur-Seine, France, \\
Laboratoire de Physique Th\'eorique et Hautes Energies (LPTHE), UMR 7589,\\ Sorbonne Universit\'e et CNRS, 4 place Jussieu, 75252 Paris Cedex 05, France}

\begin{abstract}
The gluon emission from an energetic heavy quark is suppressed in the forward direction below the angle $\Theta\lesssim m_Q/E$ for a quark of mass $m_Q$ and energy $E$ according to perturbative Quantum Chromodynamics (QCD) (``dead cone"). Another consequence is the suppression of energetic particles in the jet, which has been observed already for c- and b-quark jets. The suppression of the forward particles can be explained by an application of the Modified Leading Logarithmic Approximation (MLLA) of perturbative QCD. In this paper we investigate whether this type of analysis can be carried out also for top-quark jets with a much higher heavy quark mass allowing for QCD tests in this new kinematic regime. The new aspect of this analysis is the finite lifetime of the top-quark. We consider for simplicity the decay $t\to b\ell\nu$, where the b-quark radiates gluons as well and partially obscures the dead cone. Guided by the decay amplitude in leading order in $\alpha_s$ for $e^+e^- \to t \bar t$ we propose a method to separate the radiation by the $\widehat{tb}$ dipole in the decay process, which is superimposed on the primary radiation from the $\widehat{t \bar t}$ dipole involving the top-quark dead cone effect. The momentum distributions of partons or hadrons are determined for finite decay angles of the b-quark $\Theta_b$ and extrapolated in the forward direction $\Theta_b=0$ where the radiation from the decay process is expected to vanish. This method is successfully tested at the parton level, and the results obtained for hadrons are compatible with the MLLA relation with an accuracy of around 15\%. Our calculations are carried out with the \textsc{Pythia} 8.3 Monte Carlo Event Generator.
\end{abstract}

\maketitle

\section{Introduction}
\label{sec:intro}

Gluon emission from a coloured particle is an elementary process in QCD and it can be seen in analogy to photon emission from an electrically charged particle in QED. This radiation has a characteristic dependence on the mass of the primary particle that has been studied with its observable consequences in perturbative QCD~\cite{Dokshitzer:1991fc,Dokshitzer:1991fd}. For an energetic heavy quark $Q$ of mass $m_Q$ and energy $E_Q$, such that $E_Q/m_Q \gg 1$, the probability of gluon emission at low emission angle $\Theta_g$ and low energy $\omega$, can be written in leading perturbative order as
\begin{equation}
  d\sigma_{Q\to Q+g} \simeq \frac{\alpha_s}{\pi}C_F\frac{\Theta^2d\Theta^2}{(\Theta^2+\Theta^2_0)^2} \frac{d\omega}{\omega},
\label{emission}
\end{equation}
with an angular cut-off $\Theta_0=m_Q/p_Q\simeq m_Q/E_Q$ at momentum $p_Q$; $\alpha_s$ denotes the strong coupling constant and $C_F$ the QCD colour factor at the branching vertex $Q\to Q+g$. Therefore, for smaller emission angles $\Theta\lesssim\Theta_0$, gluon radiation is suppressed and vanishes in the forward direction.  The depopulated region around the flight direction of the heavy quark is called ``dead cone''. This depopulation increases with increasing mass. On the other hand, for large emission angles $\Theta\gg \Theta_0$, the gluon radiation pattern becomes identical to that of a light quark jet.

The dead cone effect has first been observed experimentally by the relative suppression of total particle multiplicity associated with heavy quark (c- and b- quark) production in $e^+e^-$ annihilation already some time ago~\cite{Schumm:1992xt,Dokshitzer:2005ri}. 
Only recently has the dead cone effect been established in the hadronic final state  by the ALICE collaboration~\cite{ALICE:2021aqk}. The ALICE collaboration analyzed charm-quark jets produced in pp collisions at the Large Hadron Collider and measured the angular dependence of the production of subjets with respect to the direction of the primary charm quark. They found a depletion in the forward direction as expected from QCD based Monte Carlo Event Generators (MCEG). 
Alternatively, the dead cone effect can be observed with momentum spectra of partons and hadrons which are expected to be suppressed at high momenta. This suppression has been investigated using an analytical approach within the Modified Leading Logarithmic Approximation (MLLA)~\cite{Dokshitzer:1991fd}. Recently, this suppression of high momenta predicted for heavy quark jets has been observed in $e^+e^-$-annihilation at the LEP energy of 91.2~GeV, and these results support the QCD expectations within MLLA; in particular, the dependence on the  mass of the primary heavy quark has been confirmed by the comparison of charm-quark and bottom quark-jets~\cite{Kluth:2023umf}.\footnote{
After completion of this work, a study of the dead cone effect in b-quark jets based on the angular distribution of subjets has been presented by the CMS collaboration~\cite{CMS-PAS-SMP-25-008}.
} 

In this paper, we want to investigate how the analysis of the momentum spectra could be applied to the production of top-quark jets. This would extend the dead cone study to the much higher mass of the top-quark $m_t= 172.52\pm0.33$~GeV~\cite{ParticleDataGroup:2024cfk,ATLAS:2012aj,CMS:2023tmass,ATLAS2025} 
as a test of the underlying QCD theory but also of our understanding of the hadronization effects in the presence of a large forward empty cone. A new feature in top-quark production is the finite lifetime corresponding to the width $\Gamma_t=1.42^{+0.19}_{-0.15}$~GeV~\cite{ParticleDataGroup:2024cfk}. There are two aspects to be taken into account: first, in the decays $t\to bW$ with $W \to \ell \nu$, which we consider for simplicity, the coloured quark $b$ appears from the decay in the final state which radiates itself and this radiation is superimposed over the top-quark Bremsstrahlung. Secondly, the finite lifetime will lead to a suppression of soft particles in the jet with energies of the order $\Gamma_t$~\cite{Jikia:1991yd,Orr:1992uv}.  

It is therefore our primary aim to investigate the possibilities of how to separate the top-quark gluon radiation from the ``background" of radiation from secondary decays. Two methods have already been considered: one selects decay configurations of $(t\to tb)$ with laterally emitted b-quarks, so forward radiation $(t\to b+X)$ can be separated. Secondly, from the measurement of the mass $M(b\ell \nu)$, one can decide whether the emitted gluon comes from primary radiation with final top-quark mass at $M(b\ell \nu) \sim m_t$ or from the radiation from the b-quark in top-quark decay with $M(b\ell \nu) < m_t$. Such feasibility studies based on an angular analysis of subjet production have been performed by Maltoni, Selvaggi and Thaler~\cite{MaltoniSelvaggiThaler:2016DeadCone} with the result based on MCEG that a dead cone may be made visible in the high energy top-quark jet with the application of the two methods. 

In this paper, we investigate again the potential of angular analysis with a focus on the phenomenological appearance of the decay process $t\to bW$ superimposed on the gluon emission process in $t\to tg$. However, our main aim is the study of the feasibility of observing the dead cone effect from the suppression of the high momenta in the top-quark fragmentation. Such a study is less dependent on the determination of the initial top-quark direction as required in the angular analysis. This study is based mainly on the \textsc{Pythia}~8.3 MCEG~\cite{Bierlich:2022pyt,Sjostrand:2014zea}. Event generation is performed using the default configuration, in which matrix-element corrections 
are enabled; these corrections significantly improve the description of gluon emission off heavy quarks and are important for an accurate modeling of the dead-cone effect~\cite{Alwall:2014hca}. 
Subsequently, a parton cascade with a final transition to hadrons is generated according to the Lund hadronization approach~\cite{Andersson:1983ia}. In addition we want to follow the main arguments also within the simplified treatment based on leading order perturbation theory, especially~\cite{Orr:1992uv}, and also on the results from the MLLA~\cite{Dokshitzer1991Basics} with the important concept of ``angular ordering"~\cite{Ermolaev:1981cm,Mueller:1981ex} built in, which provides some analytical results on the jet evolution. In Section~\ref{sec:theory}, we discuss some of the theoretical tools and results used in this analysis. In section~\ref{sec:angularstructure} we investigate the angular structure of top-quark  production events and study how well the top-quark radiation can be separated from the other processes, whereby we also address some aspects of hadronization. This investigation will motivate our procedure in the study of momentum distributions and their separation from ``background".

The momentum distributions obtained in this way will be compared to the MLLA expectations, thereby following and extending our previous analysis of c- and b-quark jet production~\cite{Kluth:2023umf} (Section~\ref{sec:momenta}). For this analysis, we apply the \textsc{Pythia} 8.3 MCEG~\cite{Bierlich:2022pyt} in its standard version. 
For hadrons we include charged and neutral particles as provided by the MC.
The consequences of the finite width $\Gamma_t$ of the top-quark on the soft particle production 
are investigated based on a modified version of \textsc{Pythia} 8.3 in Section~\ref{sec:finite_width}. 
We consider here the reaction $e^+e^- \to t \bar t$ at the c.m.s.\ energy $\sqrt{s}=1$~TeV which may become available in a future collider. Some aspects of the application to top-quark jets in pp collisions are considered in Section~\ref{sec:pp}.

\section{Theoretical Considerations}
\label{sec:theory}

\subsection{Gluon emission from unstable top-quark in leading perturbative order}

We begin with a description of some essential features of particle jet formation in the leading order of $\alpha_s$. In this approximation we consider the reaction $e^+e^-\to t \bar t g$  where the main top-quark decay is $t\to bW$ and we take into account here for simplicity only the leptonic decay channel $t\to be\nu$. For this decay, the gluon energy and the angular distribution have been calculated in the leading order of $\alpha_s$ \cite{Orr:1992uv,Jikia:1991yd} and the results can be written as~\cite{Orr:1992uv}
\begin{gather} 
\frac{1}{\sigma_0} \frac{d\sigma_g}{d\omega\ d\cos\theta_g d\phi_g} = \frac{\alpha_s C_F}{4\pi^2} \omega F, \label{FKhozeOtt} \\
F=\lvert A\rvert^2 + \  \lvert B_1\rvert^2 \ + \   \lvert B_2\rvert^2 \ - 2Re[B_1B_2^*] \ +  2Re[A(B_1-B_2)^*]. \nonumber
\end{gather}
with
\begin{gather}
    |A|^2=-\frac{m_t^2}{(p_t\cdot k)^2} - \frac{m_t^2}{(p_{\bar t}\cdot k)^2} + \frac{2p_t\cdot p_{\bar t}}{p_t\cdot k \ p_{\bar t}\cdot k} \nonumber\\
    |B_1|^2=-\frac{m_b^2}{(p_b\cdot k)^2} - \frac{m_t^2}{(p_t\cdot k)^2} + \frac{2p_b\cdot p_t}{p_b\cdot k \ p_t\cdot k}\nonumber\\
    |B_2|^2=-\frac{m_b^2}{(p_{\bar b}\cdot k)^2} - \frac{m_t^2}{(p_{\bar t}\cdot k)^2} + \frac{2p_{\bar b}\cdot p_{\bar t}}{p_{\bar b}\cdot k \ p_{\bar t}\cdot k} \nonumber
\end{gather}

\begin{figure} [htp!]
    \centering
    \includegraphics[width=0.8\linewidth]{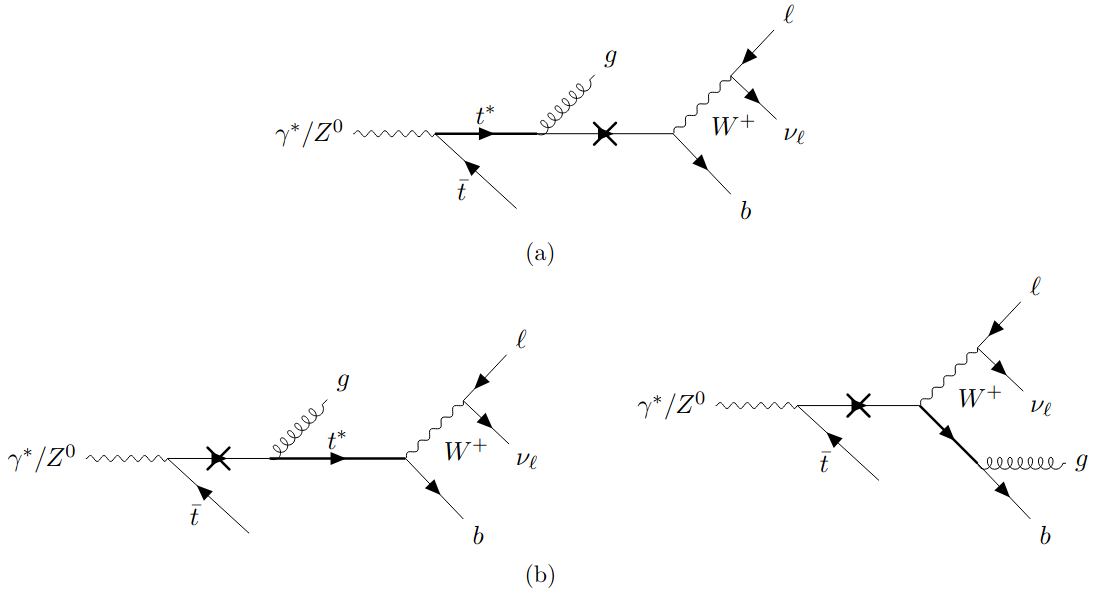}
    \caption{Feynman diagrams for gluon emission in \(e^+e^-\to t \bar t \to bW^+\bar b W^-\) relevant to the top-quark hemisphere according to the decomposition Eq.~\eqref{FKhozeOtt}. Heavy lines denote off-shell quarks. Panel~(a) shows the correction to top-quark production (amplitude \(A\)), while panel~(b) shows the corresponding corrections to top-quark decay (amplitude \(B_1\)), including gluon emission from the \(b\)-quark in the decay (according to~\cite{Orr:1992uv}).}
    \label{fig:Feynman}
\end{figure}

The amplitudes $A$, $B_1$ and $B_2$ are expressed in terms of the external particle 4-momenta of the quarks $p_i$ and the gluon $k$ and correspond to the ``dipoles"~\cite{Dokshitzer1991Basics} $\widehat{t\bar t}$, $\widehat{tb}$ and $\widehat{\bar t\bar b}$. The term with $\lvert A\rvert^2$ corresponds to the emission of a gluon from the top-quark as in Eq.~\eqref{emission}, whereas the term $|B_1|^2$ represents the decay of the top-quark $t\to b W$ with associated gluon emission from the $t$- or $b$-quark, and accordingly the term $|B_2|^2$ for the $\bar t$ decay. The corresponding Feynman diagrams relevant for the top-quark production and decay in the forward hemisphere are shown in Fig.~\ref{fig:Feynman}.

The two interference terms in Eq.~\eqref{FKhozeOtt} contain factors $D_i$ depending on the top-quark width $\Gamma_t$:
\begin{equation}
    m_t^2 D_{1}=\frac{m_t^2\Gamma_t^2}{(p_t \cdot k)^2\ +\ m_t^2\Gamma_t^2}
\end{equation}
for the t-quark and likewise for $D_2$ with the $\bar t$-quark.
As discussed in Ref.~\cite{Orr:1992uv} it depends on the particular orientation of the gluon with respect to the $t$- and $b$-quarks whether the dependence of the cross section on the width $\Gamma_t$ becomes observable. The term $2Re[B_1B_2^*]$ is relevant in the $t\bar t$ threshold region~\cite{Khoze:1994fu}. In \textsc{Pythia} 8.3 the width dependent terms are not included in the default version. They can be included within a special option and their effect is investigated in Sect.~\ref{sec:finite_width} with the result that for the application on momentum spectra of hadrons the effect is rather small of O(10\%). So in the following sections~\ref{sec:angularstructure} and~\ref{sec:momenta} we do not consider these effects.

In the following we study the results for the hemisphere of the top-quark only which is also relevant for the application to $pp$-collisions. 
The above results in Eq.~\eqref{FKhozeOtt} can be written for the top-quark hemisphere in the approximation for massless gluons with small gluon angle $\Theta_g$ to the top-quark and $\Theta_g'$ to the b-quark at high primary energies $E_t$ as
\begin{alignat} {1}
  \omega^2 |A|^2&= \frac{4\Theta_g^2}{(\Theta_g^2+\Theta_0^2)^2} + O(\Theta_0^2) \label{termA} \\
  \omega^2 |B_1|^2&=  \frac{8(1-\beta\beta'\cos\gamma)}{(\Theta_g^2+\Theta_0^2)(\Theta_g'^2+\Theta_0'^2)}-
  \frac{4 \Theta_0^2}{(\Theta_g^2+\Theta_0^2)^2} - \frac{4 \Theta_0'^2}{(\Theta_g'^2+\Theta_0'^2)^2}+ O(\Theta_0^4,\Theta_0'^4) \label{termB}
\end{alignat}
where $\beta,\beta'$ denote the t- and b-quark velocities $\beta=p_t/E_t$, $\beta\approx 1-\frac{1}{2} Q_0^2$ and $\beta'=p_b/E_b$, $\beta'\approx 1-\frac{1}{2} Q_0'^2$ with $\Theta_0'=m_b/p_b$; $\gamma$ denotes the decay angle of the b-quark with respect to the t-quark direction in the $e^+e^-$ c.m.s.\ and $p_i$ denotes the 3-momenta.

In the leading order calculation considered here, we note that the b-quark radiation comes from the $\widehat{tb}$ dipole and therefore, the b-quark radiation is associated with additional radiation from the top-quark. Then, even if the b-quark decays with large angle~$\gamma$ such that the b-quark fragments are well separated from the top-quark fragments, there remains some radiation in the top-quark direction $( \propto 1/(\Theta_g^2+\Theta_0^2))$ from the $\widehat{tb}$ dipole, which is represented by the first term in Eq.~\eqref{termB} and is finite at $\Theta_g=0$; the two other terms vanish at high energies. However, we observe that the leading term in $|B_1|^2$ vanishes at high energies for decays with the b-quark in the forward direction $\gamma=0$. This also holds for the gluon momentum spectrum after integration over the gluon decay angles $\Theta_g, \Theta_g'$.

In order to obtain the momentum spectra of partons from top-quark fragmentation we have to integrate these distributions over the gluon emission angles and obtain contributions from the different components. For $|A|^2$ we get at large angles $\Theta_g$ the integral over $d\Theta_g^2/\Theta_g^2$ which yields a contribution logarithmically rising with the upper limit. For the terms in $|B_1|^2$  the integrals over $\Theta_g$, $\Theta'_g$ are dominated by the small angle emissions near $\Theta_0,\Theta_0'$ and we can approximately write for the angular integral
\begin{equation}
   \omega^2 \langle|B_1|^2\rangle \simeq 8 c_1(\gamma) (1-\beta\beta'\cos\gamma) + 4c_2 \Theta_0^2  + 4c_3 \Theta_0'^2 \label{B1average}
\end{equation}
At the high energies considered here the second and third terms are negligible and so $\langle|B_1|^2\rangle$ 
decreases with decreasing decay angle $\gamma$ for $t\to b$. For our scenario with $E_t=500$~GeV we have $\beta=0.94$ and $\beta'=0.995...0.9999$ for the b-quark momenta in the region 20 - 450~GeV. While in the first term $c_1(\gamma)$ remains finite, there is also the factor $(1-0.94 \cos\gamma)$ which nearly vanishes for the emission of the b-quark parallel to the top-quark. This behaviour is well understood, and, in fact, for particles of equal mass and velocity, it follows from Eq.~\eqref{termB} 
\begin{equation}
  \omega^2 \langle|B_1|^2\rangle \simeq(1-\cos\gamma)+O(\Theta_0^2),     
\end{equation}
which vanishes for $\gamma\to 0$, as by a Lorentz boost into the rest frame the two particles at rest would clearly not radiate. On the other hand, the $|A|^2$ term remains large in the limit $\gamma\to 0$ in comparison to the $|B_1|^2$ term. We conclude that for the b-quark emitted in the top-quark decay in the forward direction the radiation from the $\widehat{tb}$ dipole nearly vanishes. This property opens the possibility to obtain separately the gluon radiation pattern of the stable top-quark represented by the $|A|^2$ term in Eq.~\eqref{FKhozeOtt}. 
As the two jets produced by the top- and the b-quark cannot be separated directly if they are parallel, we consider the possibility to determine the momentum spectra of the gluons at larger relative angles $\gamma$ and extrapolate the results so obtained to the angle $\gamma=0$. The same should also apply for the final momentum spectra of partons or hadrons as they are generated by the further evolution of the primary gluon. If the $\widehat{tb}$ dipole does not radiate gluons, neither partons nor hadrons will evolve from this source. In this way the momentum spectrum corresponding to a stable top-quark can be obtained by extrapolation. 

We note that in the leading approximations 
Eqs.~\eqref{FKhozeOtt}, \eqref{termB} the momentum spectrum factorizes from the angular distributions and the momentum spectrum changes with angle $\gamma$ only in magnitude not in shape, but this does not hold in general. For parton and hadron spectra, besides the $\gamma$ dependence in 
Eq.~\eqref{B1average} there is also a change of the dipole radiation with this angle $\gamma$ as a consequence of the jet evolution. 
The extrapolation function for the parton and hadron momentum spectra will be determined here from the MCEG. 

\subsection{Multiple gluon emission and angular ordering}
\label{multiplegluon}

In this subsection we will recall a few properties of jet evolution which are of particular relevance for our subsequent analysis. The primary gluon emitted from the initial quark is the source for the subsequent parton production which can be described by an evolution equation for the bulk of particles within the Leading and Modified Leading Logarithmic Approximation (DLA and MLLA) in a one loop order approach. For illustration we recall the evolution equation in DLA for the distribution $D_q^p$ of partons $p$ with energy $\omega$ in a jet of opening angle $\Theta$ for a primary quark $q$ of energy $E_q$~\cite{Dokshitzer1991Basics} 
\begin{equation}
    D_q^p(\xi,y_\Theta)= \delta_q^p\delta(\xi) \ + \ \int_0^\xi d\xi' \int_0^{y_\Theta} dy'\ (C_F/N_c)\ \gamma_0^2(y')\ D_g^p(\xi',y')
    \label{evolutioneq}
\end{equation}
with $\xi=\ln(E_q/\omega)$, $y_\Theta= \ln(\omega\Theta/Q_0)$, $\gamma_0^2=4N_c/(b\ln y')$, $b=\frac{11}{3} N_c-\frac{2}{3}n_f$ and cut off parameter $Q_0$. Because of ``angular ordering"~\cite{Ermolaev:1981cm,Mueller:1981ex} the secondary particles are emitted within the cone given by the production angle of the primary gluon with parameters $\xi',y'$, the production of particles outside this cone vanishes in the azimuthal average\footnote{For an introductory outline of intra-jet coherence phenomena, see for example Refs.~\cite{Dokshitzer1991Basics,Ellis:1996mzs}.}. For a primary heavy quark the modified angular distribution as in Eq.~\eqref{emission} has to be inserted under the integral above for the first emission. 

\begin{figure} [t]
    \centering
     \includegraphics[width=1.01\linewidth]{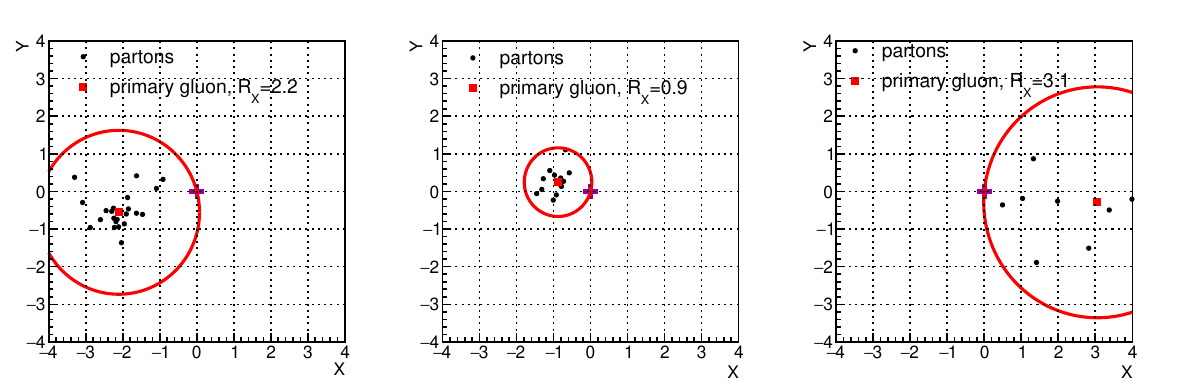}
    \includegraphics[width=1.01\linewidth]{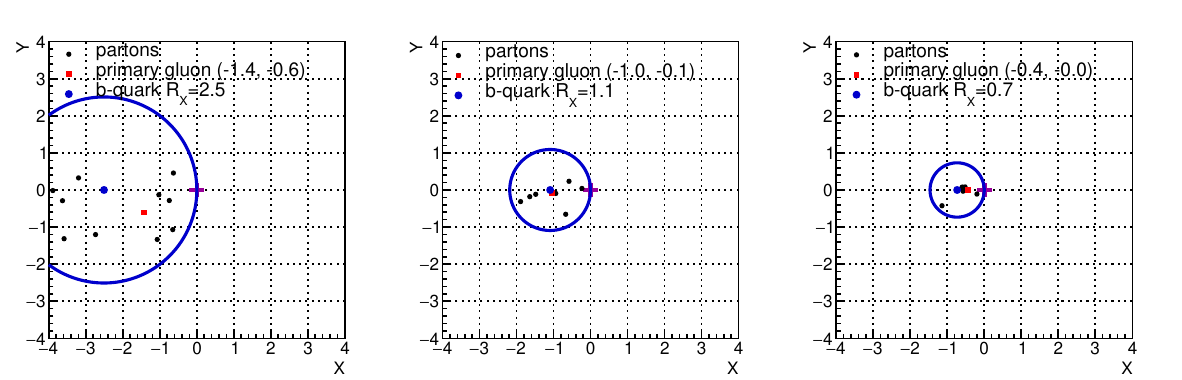}
    \caption{
    Scatter plots for three representative events from $e^+e^-\to t \bar t+X$ with stable top-quark at $\sqrt{s}=1$ TeV (upper panels), showing the distribution of partons from top-quark fragmentation over the angles $X$ and $Y$ in Eq. \eqref{XYdef} in units of the dead cone angle $\Theta_0$ with the top-quark direction in the center (0,0) (shown as cross): the primary gluon as red point at angle $R_X=\Theta/\Theta_0$ to the centre surrounded by the emitted partons (black points) limited by the same angle $R_X$ (red circle) as required by ``angular ordering"; lower panels:
    Scatter plots as above but for events with top-quark decay $t\to b\ell \nu +  g$ for primary gluons triggered with $X_g<0$ shown as red square and the b-quark as blue dot rotated into negative $X$-direction  ($Y=0$) and with $X_b<-0.5$. The emitted partons come mainly from the b-quark and are found inside the blue circle around the b-quark position as required by angular ordering; here $\Theta_0=20^\circ$ at 1~TeV (from \textsc{Pythia} 8.3). }
    \label{fig:t-b-circles}
\end{figure}

As an illustration of this production process with angular ordering we show in the upper panels of Fig.~\ref{fig:t-b-circles} the distribution of the final partons over the rescaled angular variables $(X,Y)$
for a few typical events obtained by the \textsc{Pythia} 8.3 MCEG of the process $e^+e^- \to t \bar t$ at 1~TeV for stable top-quarks with~\cite{MaltoniSelvaggiThaler:2016DeadCone}
\begin{equation}
  X=(\Theta/\Theta_0) \cos \phi, \qquad   Y=(\Theta/\Theta_0) \sin \phi.
\label{XYdef}
\end{equation}     
Here the production angle $\Theta$ of a parton is measured with respect to the top-quark direction in the center at (0,0) and is normalized to the dead cone angle $\Theta_0=m_t/E_t$, and $\phi$ is the azimuth around the top-quark direction with $\phi=0$ for the initial $e^-$ as defined in the App.~\ref{app:kinematics}. The production angle of a primary gluon is indicated by the red point, then the partons emitted from this gluon, because of angular ordering, fall inside the circle around the gluon whose radius is defined by the distance to the top-quark direction at $(0,0)$ (the red circle through the origin). 

In the realistic case of an unstable top-quark we have in addition to the term $|A|^2$ for the $t \bar t$ antenna in Eq.~\eqref{FKhozeOtt} also terms $|B_{1,2}|^2$ with the gluon emission from the $b$- and $\bar b$-quarks. In this leading order approximation in $\alpha_s$ the primary gluon is radiated from either one of the superimposed dipoles and therefore 
it will be produced either from the top- or from the b-quark. As a consequence, individual events should show dominantly radiation from either source, t- or b-quark. 

This situation can be visualized again in the ($X,Y$) angular scatterplot of the final partons. We orient the event in such a way that the b-quark moves in the direction of the negative $X$-axis.  Then, if we select a parton with $X>0$ it will primarily come from the primary gluon emitted from the top-quark and the parton distribution will look similar to the event in the rightmost panel of the upper row.

In the lower panels of Fig.~\ref{fig:t-b-circles} we show representative events  
with primary gluons found in the left hemisphere with $X_g<0$. They can be related to the emission from a b-quark centered along the negative $X_b$ axis with $X_b<-0.5$. The partons are found within a circle around the b-quark limited by the direction of the top-quark at (0,0) as required by angular ordering in the radiation from the $t\to b$ transition. While in these events all partons are confined into the blue circle there are also events with partons outside which have its origin from t- and b-quark radiation processes. 

These figures show the important role played by the angular ordering condition: without this restriction b-quarks at small angles $X_b$ would emit a large fraction of partons into the right hemisphere. With this condition fulfilled, it is of advantage to determine the top-quark fragmentation products only from particles in the right hemisphere $X>0$ where the b-quark jet radiation is suppressed.

\section{Angular structure of top-quark jet}
\label{sec:angularstructure}

\noindent
In this section, we study the angular distribution of partons and hadrons in the top-quark jet with the aim to reveal the main elements of the leading order description of Eq.~\eqref{FKhozeOtt} from a jet analysis, that is to confirm the leading role of the dead cone amplitude $A$ and to identify the $t\to bW$ decay amplitude $B_1$ (the ``background") that causes a modification of the dead cone phenomenology. Restricting to the leptonic decays $W\to e \nu$ for simplicity, the final state after top-quark decay includes in the lowest order a b-jet and a gluon jet; for a stable top-quark only the top-quark jet and a gluon jet would appear in the final state. These investigations are important for the design of our subsequent study of momentum distributions. The results are obtained for the top-quark hemisphere in the primary reaction $e^+e^- \to t \bar t$ at the c.m.s.\ energy $\sqrt{s}=1$~TeV using the \textsc{Pythia} 8.3 MCEG. 

In order to obtain jets from the final state of partons or hadrons, these particles are clustered using the jet reconstruction package \textsc{FastJet}~\cite{Cacciari:2011ma,Cacciari:2005hq} with the $e^+e^-$ generalized-$k_t$ algorithm in inclusive mode, choosing parameters $p=1$ and radius $R=0.2$. The relatively small radius is selected to keep the b-jet core well separated from the gluonic jets originating from the top-quark. All jets with the exclusion of the b-quark jets are combined into a single jet by adding their 4-vectors to represent the primary gluon in the lowest order formula Eq.~\eqref{FKhozeOtt} which is emitted either from the top-quark or from the b-quark. This choice of radius $R$ is consistent with the radii commonly used in heavy-flavour and b-jet studies~\cite{ATLAS:2016bjetSubstructure,ATLAS:2013BoostedTop,CMS:2013bjetPerf}. 
A larger radius like $R=0.4$ would not separate the b-jet and the gluon jet in a substantial fraction of events, especially if the g-jet and b-jet both originate from top-quark decays. At the generator level $g\to b\bar b$ splittings also occur, but they are rare in the selected sample and occur only in about $1\%$ of the events. In our simulation of events, we choose for the b-jet the b-quark from the $t\to Wb$ decay chain. 

We have also investigated alternative jet definitions within the $(e^+e^-)$ generalized-$k_t$ options, in particular $p=0$ (Cambridge/Aachen) and $p=-1$ (anti-$k_t$), and find that the reconstructed jets and the observables used in this analysis show only a mild sensitivity to the choice of $p$, shifting the momentum distributions by at most $5\%$. The main results remain unchanged, and therefore we adopt $p=1$ as the baseline configuration throughout.

In this analysis, the top-quark direction is known from its construction within the MCEG. 
In the experiment the determination of the top-quark direction is limited not only by fluctuations of final-state hadrons in the jet, but also by the uncertainty in the reconstructed neutrino direction~\cite{Martin-Ramiro:2020yez,Behnke:2013lya}. For the di-lepton final state a resolution for the neutrino 4-vector of less than 5\% for 80\% of the events is predicted~\cite{Martin-Ramiro:2020yez}. 
Therefore, results reported on angular distributions can be compared with those of the experiment only after the corresponding averaging. 

\begin{figure}[tpb]
  \centering
  \begin{minipage}[t]{0.49\linewidth}
    \centering
    \includegraphics[width=\linewidth]{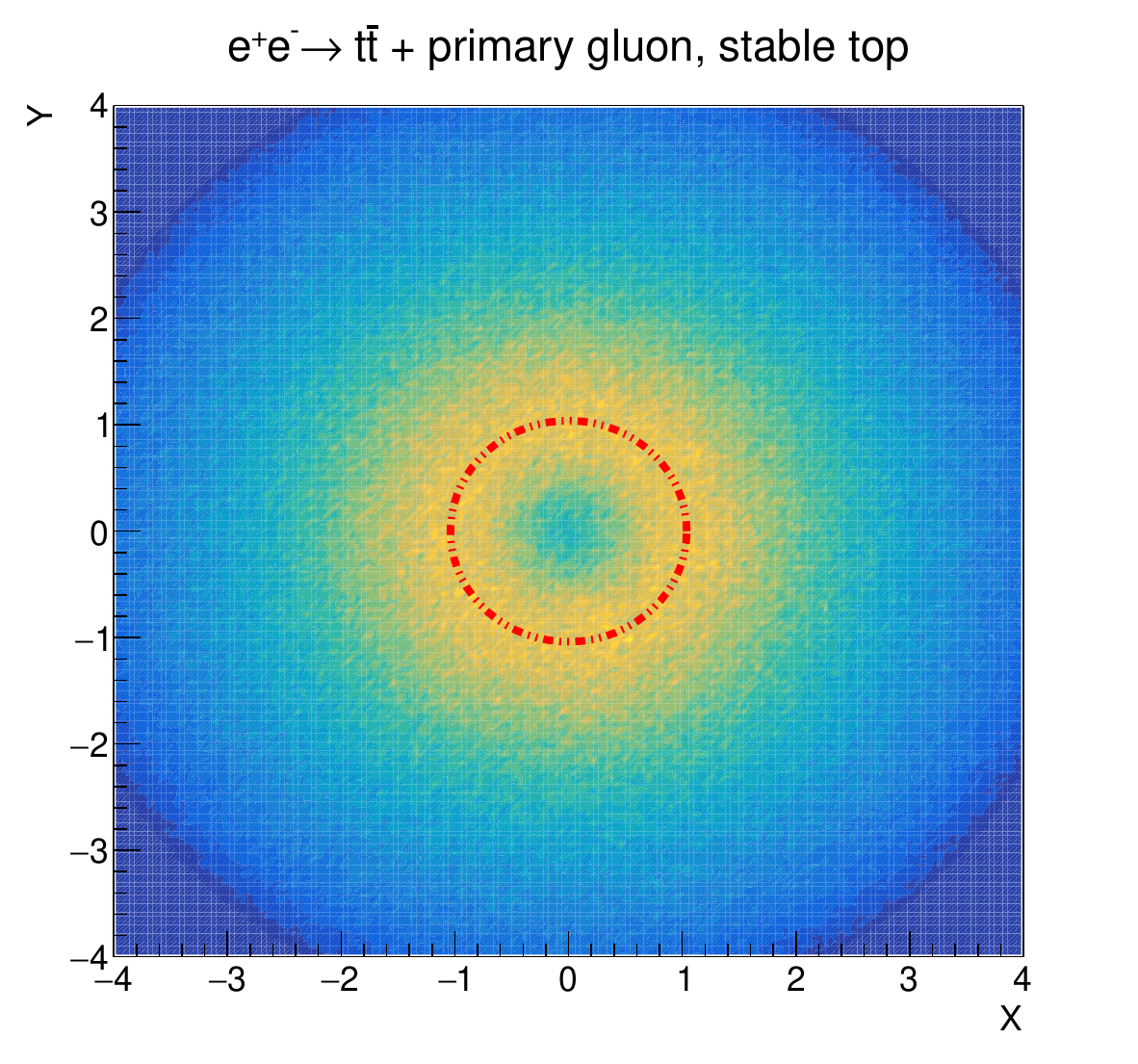}
  \end{minipage}\hfill
  \begin{minipage}[t]{0.49\linewidth}
    \centering
    \includegraphics[width=\linewidth]{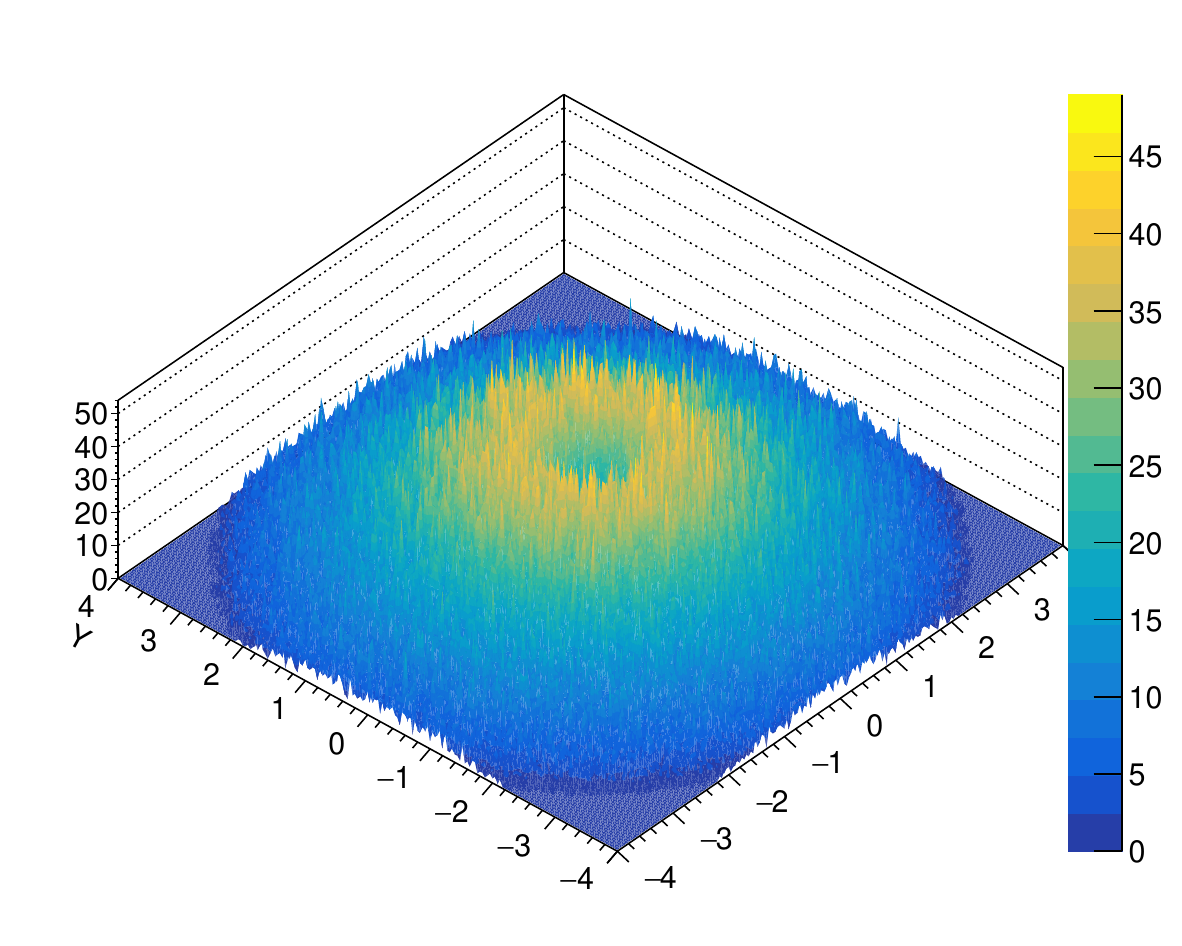}
  \end{minipage}
  \caption{Angular distribution of the reconstructed primary gluon in $e^+e^- \to t\bar t \ +$ partons at 1~TeV around the top-quark direction for stable top-quarks: the left panel shows the distribution in the rescaled angles $(X,Y)$, defined in 
  Eq.~\eqref{XYdef}, with the dead cone boundary $R_X^2=X^2+Y^2 \simeq 1$ and the suppression in the centre, the right panel shows the corresponding gluon density. At this energy the dead cone angle $\Theta_0 \simeq 20^\circ$ and the forward hemisphere ends at $R_X\simeq 4.5$. Data generated with \textsc{Pythia} 8.3.}
  \label{fig:ring}
\end{figure}

\begin{figure}[t]
\begin{tabular}{cc}
\includegraphics[height=8.5cm,width=9.0cm]{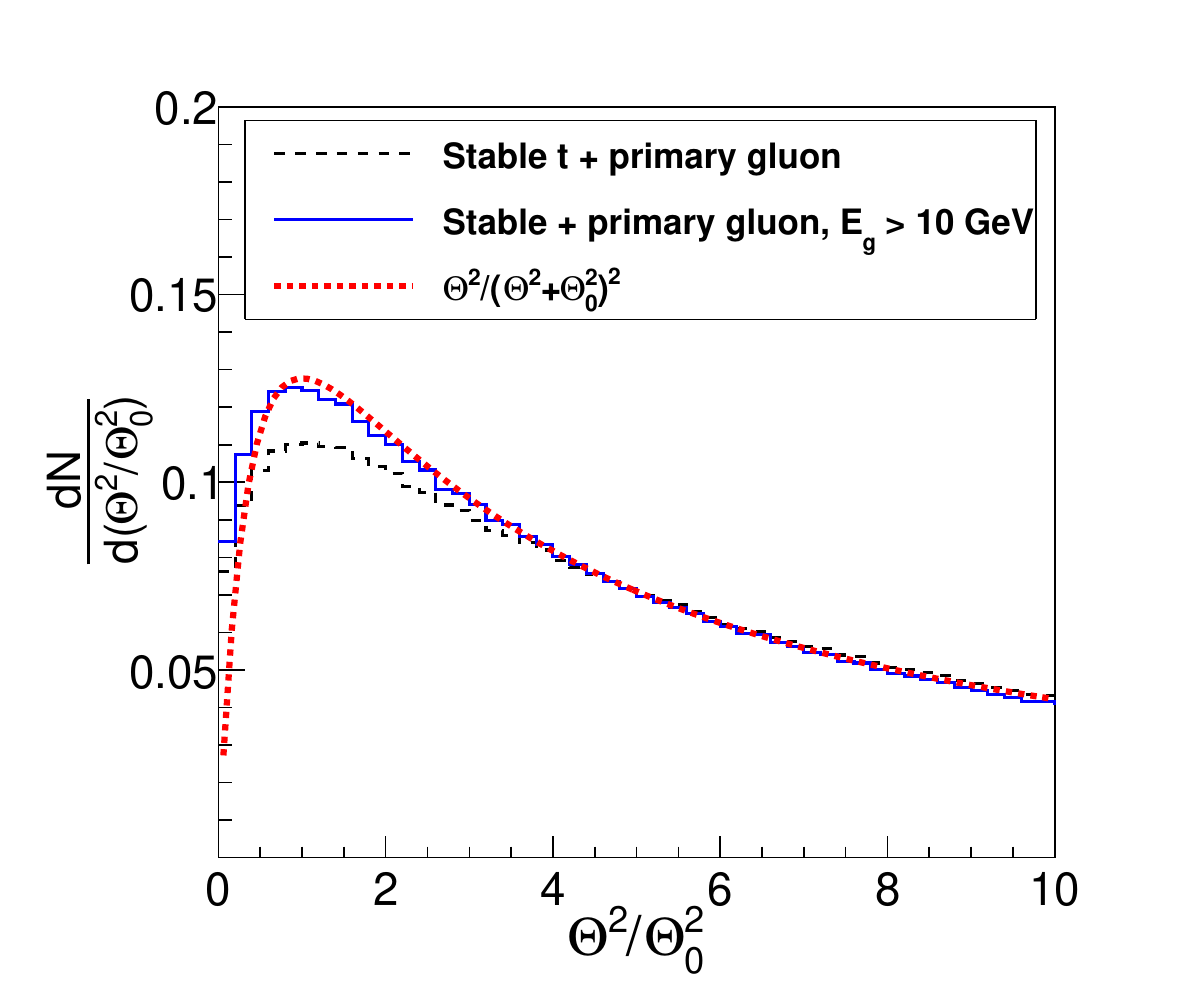} &
\end{tabular}
\begin{tabular}{cc}
\hskip -1.2cm
\includegraphics[height=8.5cm,width=9.0cm]{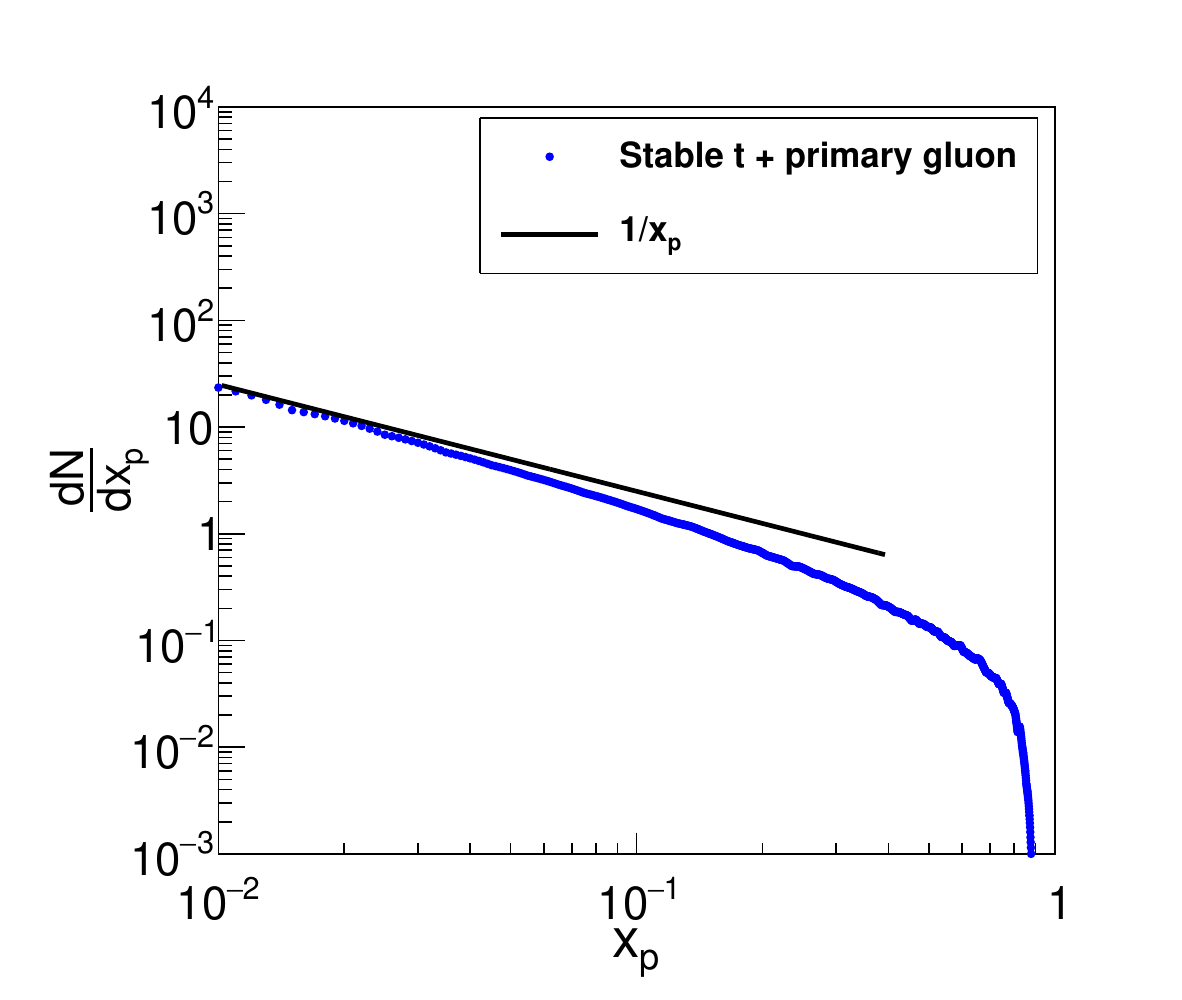} 
\end{tabular}
\caption{Distribution of the reconstructed primary gluon as in Fig.~\ref{fig:ring} in the rescaled opening angle $\Theta/\Theta_0$ for stable top-quark production without and with gluon energy cut-off $E_g>10$~GeV and distribution in momentum fraction $x_p$, both in comparison with the approximate analytic form Eq.~\eqref{emission}.}
\label{fig:Theta2}
\end{figure}

\subsection{Stable top-quark}

We begin with an analysis for the case of a hypothetical stable top-quark which can be treated with the methods discussed before~\cite{Kluth:2023umf}. The results serve as a benchmark in the study of unstable top-quarks to test the validity of the background subtraction from the decay processes. In the soft and collinear approximation to the radiation process the emission of a gluon by the stable top-quark is given in leading perturbative order by the amplitude $A$ in Eq.~\eqref{FKhozeOtt} corresponding to the distribution in angle and energy given in Eq.~\eqref{emission} with the dead cone angle $\Theta_0 \simeq m_t/E_t$. 
As discussed above, we first construct the jets for the radius $R=0.2$ from the parton final state using \textsc{FastJet}, then all jets except for the top-quark jet are combined into a single jet, taken as the primary gluon jet corresponding to the leading order process as for example in Eq. \eqref{evolutioneq}.

In Fig.~\ref{fig:ring} we show the 2-dimensional angular distribution of these primary gluon jets
in the rescaled angular variables $(X,Y)$ from Eq.~\eqref{XYdef} in the top-quark hemisphere of the  $e^+e^- \to t \bar t g$ process. The distribution clearly shows the suppression in the center $(X,Y)=(0,0)$ corresponding to the top-quark flight direction.  

In Fig.~\ref{fig:Theta2}, left panel, we show the distribution of the reconstructed
primary gluon jet in the polar angle $\Theta$ as dashed line together with the analytic form Eq.~\eqref{emission}, proportional to $\Theta^2/(\Theta^2+\Theta_0^2)^2$, both distributions normalized to unity over the full angular region. There is a deviation at small angles, but this disappears if we restrict the energy of the primary gluon to $E_g>10$ GeV. We relate this low energy effect to the edge of the top-quark hemisphere where gluon jets  are reconstructed from partons in an incomplete way.
Since these effects are small, we do not treat them separately in
the following. The distribution of the fractional momentum $x_p=p_g/E_t$ shown in
the right panel follows $dN/dx_p \propto 1/x_p$ as in Eq.~\eqref{emission} in the small energy region.

We conclude that our jet analysis for the stable top-quark jet with its construction of the primary gluon  reproduces the lowest order expectation of Eq.~\eqref{emission} and the $A$ amplitude in Eq.~\eqref{FKhozeOtt}. The corresponding angular analysis for partons instead of primary gluons is shown in Fig.~\ref{fig:Theta2all} in the Appendix. The ring at $R_X\sim 1$ is still visible, but the depletion near the center at $\Theta=0$ is less pronounced compared to the primary gluons.

\subsection{Decaying top-quark, 2-dim angular distributions}

Now we turn to the realistic case of an unstable top-quark with decay width
$\Gamma_t = 1.42$ GeV. The main decay channel is $t \to b W$, where we consider
only processes with leptonic decays $W \to e\nu$ for simplicity. 
The partonic or hadronic final states are now formed by a gluon jet and a b-quark jet. 
Jets with radius $R=0.2$ are obtained again from the \textsc{FastJet} analysis as discussed above, the b-jet is defined as the one which contains the b-quark or a B-hadron, the effect from $b\bar b$ production is negligible. The primary gluon is again constructed as sum over all jets except for the b-quark jet. According to Eq.~\eqref{FKhozeOtt} this jet represents the gluon emitted in leading order either from the initial top-quark (amplitude $A$) or in the top-quark decay from the top- or from the b-quark (amplitude $B_1$). 
\begin{figure}[t]
\includegraphics[height=11.5cm,width=15.5cm]{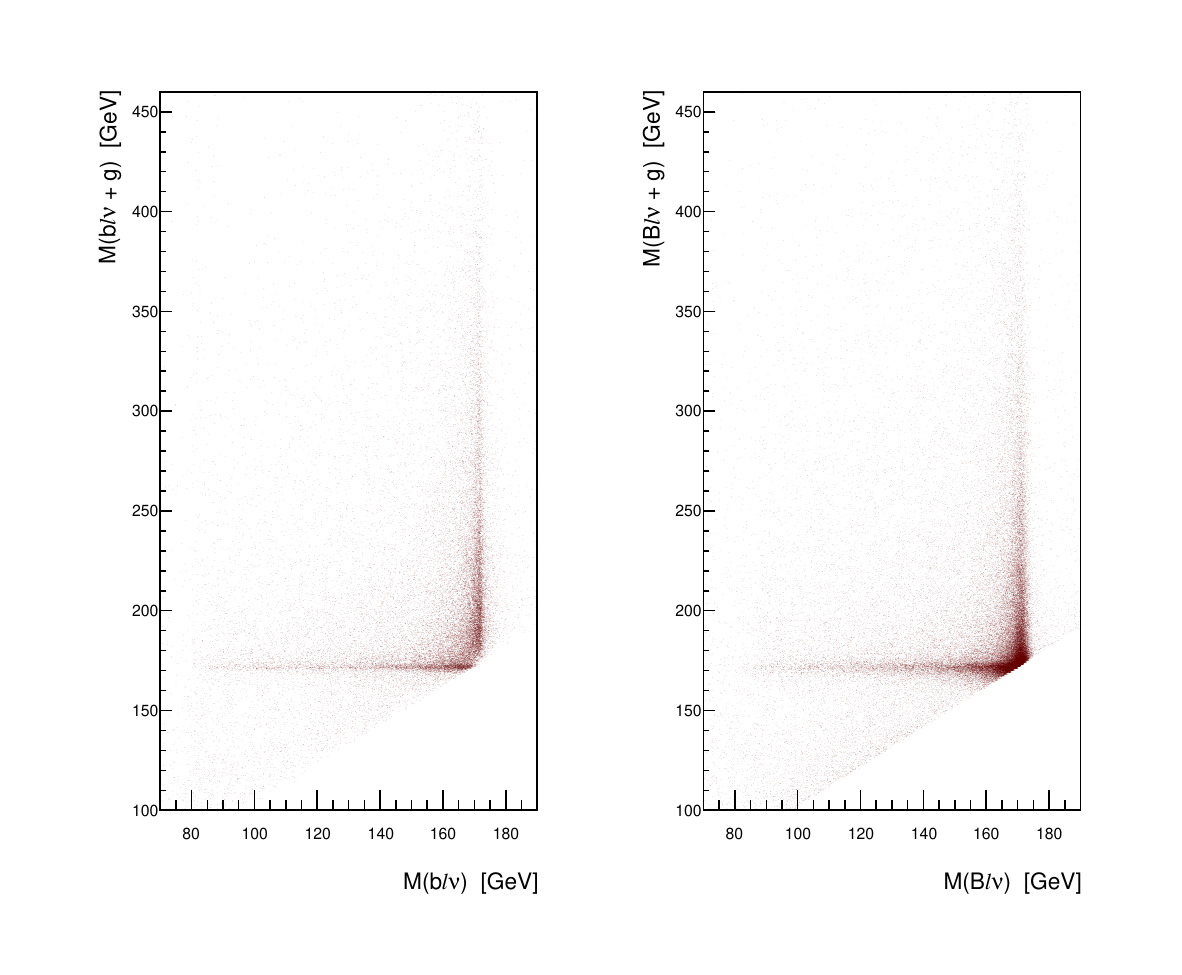} 
  \caption{The invariant masses M($b\ell \nu+ g$) vs.\ M($b\ell \nu$) in $e^+e^- \to t \bar t +X$ with $b$ and $g$ representing jets reconstructed from the final partons (left panel) and  M($B\ell \nu+ g$) vs.\ M($B\ell \nu$) from the final charged and neutral hadrons (right panel). The narrow stripes form near the top-quark mass $m_t=172.5$ GeV. The vertical stripe corresponds to the processes of hard gluon radiation from the top-quark $t^*\to tg$, the horizontal stripe to the top-quark decays with gluon emission $t\to b \ell \nu +g$. }
\label{fig:top_stripes}
\end{figure}

These two processes can be distinguished by different virtual masses as is visualized in Fig.~\ref{fig:top_stripes}. In the left panel a plot of the invariant masses M($b\ell \nu +g$) versus M($b\ell \nu$) is shown for the parton final state, in the right panel for hadrons with b-quark replaced by B-hadron. The presence of two stripes is clearly visible in both figures, the events fall with a large probability into one of the two classes. The vertical stripe of events corresponds to a fixed mass $M(b\ell \nu)$ or $M(B \ell \nu)$ near the top-quark mass and represents the events with gluon emission from the primarily produced virtual top-quark $t^* \to t g$ as in the case of stable top-quark ($\widehat{t\bar t}$ dipole in amplitude $A$). 
The horizontal band corresponds to events from the $\widehat{tb}$ dipole (amplitude $B_1$) with the invariant mass of the particles $M(b\ell\nu +g)$ or $M(B\ell\nu + g)$ near the top-quark mass corresponding to the decay $t\to b \ell\nu+g$ with variable virtual b-quark or t-quark mass.
The maximal density along the stripes is obtained at the mass scale $m_t=172.5$ GeV for the respective mass combinations, where the gluon has the minimum momentum; in this mass range, both components overlap. It is remarkable that the stripes for hadrons and partons show a rather similar density over the stripes although hadrons are generated in a later phase of the jet evolution with a higher particle multiplicity. It is essential here that hadron decays with photons like $\pi^0\to \gamma \gamma$ are included. 

There are events in Fig.~\ref{fig:top_stripes} that cannot be directly attributed to one of the two stripes. These could be events from processes with more than one gluon, such as expected in non-leading order for events with  two gluons, one from t- and one from b-quark. There could also be events where in a top-quark decay one particle is lost by leaving the hemisphere of the produced primary top-quark considered here, which would move the event outside the stripe. Such effects could contribute to a systematic error that we will estimate for our final results. 

\begin{figure}[t!]
\includegraphics[height=12cm,width=17cm]{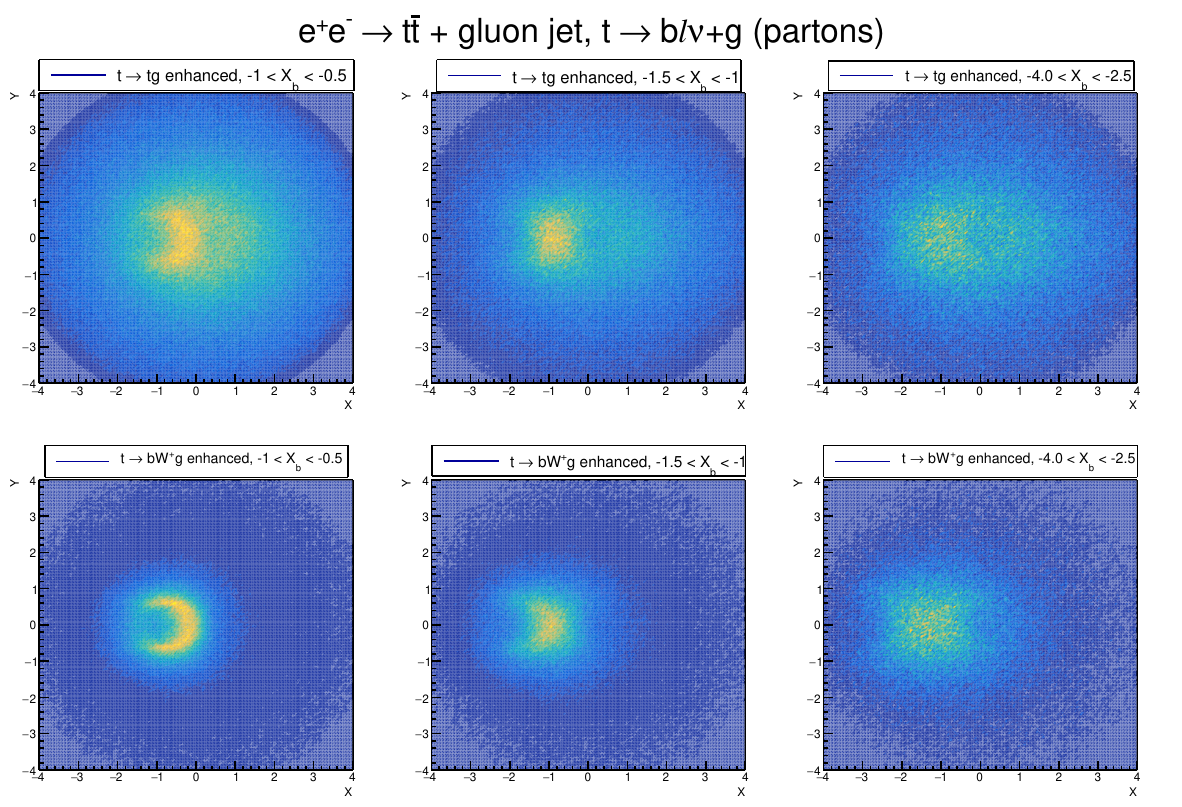} 
  \caption{Distribution of primary gluon jets constructed from partons over the angles ($X,Y$) of Eq.~\eqref{XYdef} around the top-quark at (0,0) with the b-jet pointing into the negative X-axis, for processes dominated by gluon emission $t\to tg$ (upper row) and dominated by top-quark decay $t\to b \ell \nu+ g$ (lower row) according to selection of vertical or horizontal stripes in Fig.~\ref{fig:top_stripes} 
  respectively for masses 170-175 GeV; for different angles  $X_b$ of the b-quark jets to the top-quark, increasing from left to right, b-jets removed.  The upper row reflects at low intensity the gluon radiation from the top-quark around a circle of 
  radius $R_X=1$ 
  in $X,Y$ units around the top-quark direction with the dead cone in the center, nearby on the left a yellow hot spot limited by two circles, on the left by the empty cone of the b-jet of angular radius $R=0.2$, or $R_X= 0.6$, and on the right by the circle through the origin required by angular ordering. The lower row shows the gluon radiation of the b-quark from top-quark decay outside the angular radius $R=0.2$; the wider gluon radiation from the top-quark as seen in the upper figures with $R_X\sim 1$ is suppressed. 
}
\label{fig:scatter_gluon}
\end{figure}

At first, we look again at the 2-dimensional ($X,Y$) angular distribution of the gluon jets as in the case of the stable top-quark. 
The additional radiation from the b-quark can be made more visible by selecting events with the b-quark jet emitted away from the top-quark direction. For a better presentation, the event is rotated around the top-quark direction so that the b-quark points into the negative X-axis. The results for the gluon jets are shown in Fig.~\ref{fig:scatter_gluon} for three different intervals of the angles  $X_b$ of the b-quark jet in between -4.0 and -0.5, the b-quark jets are removed from the figure. The upper row shows the events selected from the vertical stripe for masses m$(b\ell\nu)$ = 170-175~GeV, i.e.\ with $t^*\to tg$ processes enhanced, the lower row selected from the horizontal stripe with $t\to b\ell \nu + g$ processes enhanced, within the same top-quark mass interval and with m$(b\ell\nu)<160$~GeV.

In the upper part of Fig.~\ref{fig:scatter_gluon} the radiation from the top-quark around the center (0,0) with a ring of radius $\ R_X\sim 1$ with $R_X=R/\Theta_0$ in $(X,Y)$ units, i.e.\ at the dead cone angle $\Theta=\Theta_0$, is apparent at all b-quark jet angles $X_b$ and also the dead cone in the center is visible as depletion of central density, but less pronounced as in Fig.~\ref{fig:ring}. The narrow yellow stripe left from the center represents the gluon radiation from the b-quark outside the angular radius $R=0.2$ of the b-quark jet corresponding to $R_X\approx 0.6$, while the interior is removed from the figure. This additional radiation from the b-quark is limited on the right side by the circle centered at the angle $X_b$ of the b-quark jet and going through the top-quark direction at $(0.0)$ because of the angular ordering condition (see Sect.~\ref{multiplegluon} and Fig.~\ref{fig:t-b-circles}, bottom panels). This b-quark radiation decreases with increasing angle $X_b$ due to the related decrease in b-quark energy. We conclude that our selection of events from the vertical stripe in Fig.~\ref{fig:top_stripes} does not exclude the events with gluons from b-quark decays altogether. The presence of b-quark jets signals the presence of the $\widehat{tb}$ dipole with amplitude $B_1$ in Eq.~\eqref{FKhozeOtt}, which also contributes to a partial filling of the dead cone, as will be discussed further below.
 
The panels in the lower part of Fig.~\ref{fig:scatter_gluon} show the angular distribution of gluon jets for events along the horizontal stripe in Fig.~\ref{fig:top_stripes} with gluons from top-quark decays enhanced. Indeed, the dominant feature is now the radiation of half-moon like shape around the b-quark jet outside the angular radius $R=0.2$ and inside the circle through the origin $(0,0)$ required by angular ordering, with little or no evidence for the top-quark radiation around the center with radius $R_X\sim 1$. 

In a corresponding procedure we also studied the 2-dimensional ($X,Y$) distributions for final state hadrons. Two jets are constructed as before, the b-quark jet carrying the B-hadron and the primary gluon jet. The distributions of gluon jets indicate again a dead cone surrounded with a circular distribution of radius $R_X\sim 1$, while for partons it is less visible and it has disappeared for hadrons, for more details, see App.~\ref{app:angle}, Figs.~\ref{fig:figure11up.pdf} and~\ref{fig:figure8-10up.pdf}. 

\subsection{Angular distributions within a central stripe along $X$ direction and evidence for the $\widehat{tb}$ dipole}

\begin{figure}[t]
    \centering
    \includegraphics[width=0.52\linewidth]{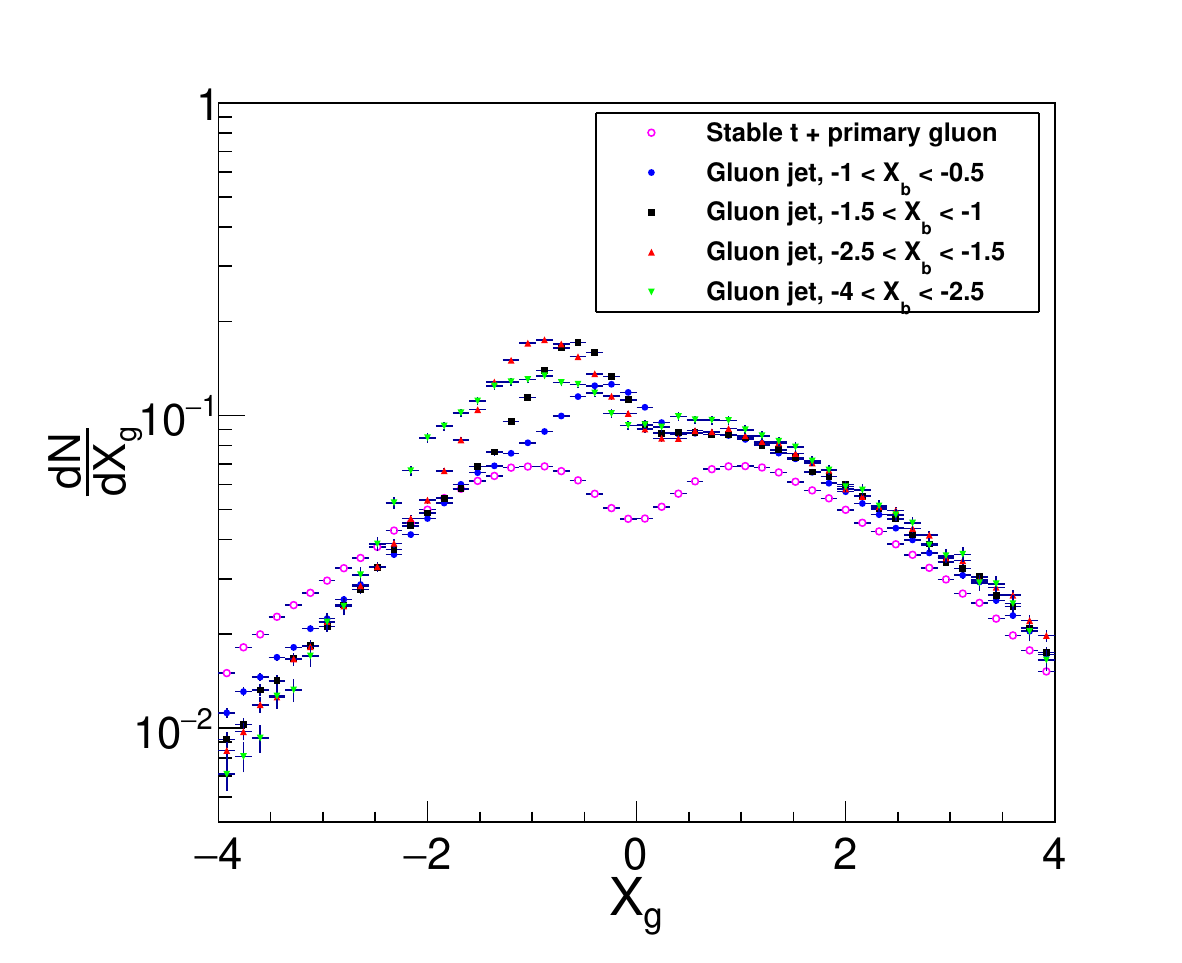}
    \hskip -1.0cm
    \includegraphics[width=0.52\linewidth]{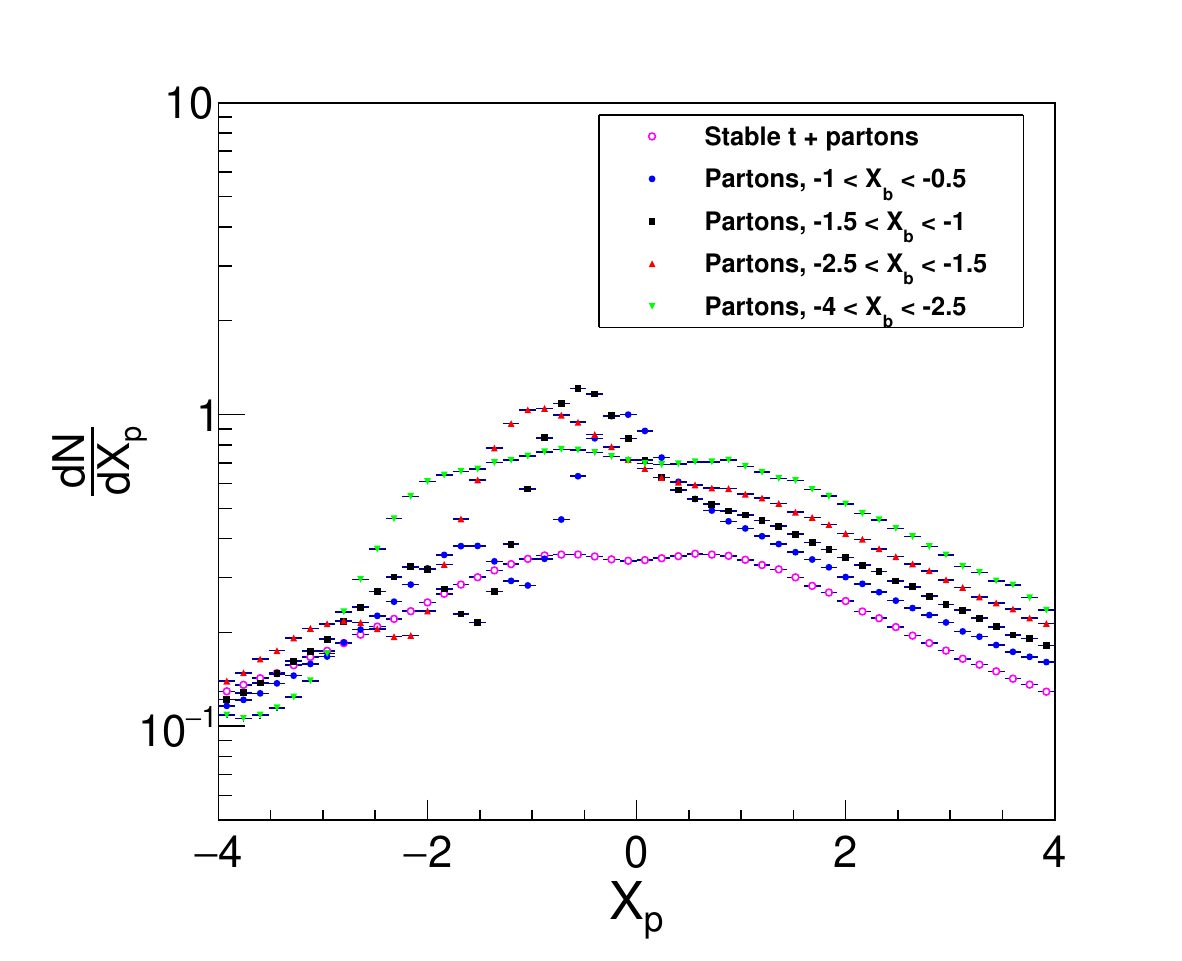}
    \caption{Distribution over angle $X_g$ of primary gluon jets, from events dominated by emission $t\to tg$, along central stripe through $(X,Y)$ scatter plot within $-\frac{1}{2} < Y < +\frac{1}{2}$ in Fig.~\ref{fig:ring} and  Fig.~\ref{fig:scatter_gluon}, upper row, for different regions in the b-quark jet angle~$X_b$. The radiation from the $\widehat{tb}$ dipole is superimposed to the radiation from the stable top-quark and is strongest in top-quark direction at $X_g=0$ (left panel); the corresponding distribution for partons in the same central stripe show a stronger separation of distributions for $X_g\gtrsim 0$ with angle $X_b$, obtained from Fig.~\ref{fig:Theta2all} and Fig.~\ref{fig:figure8-10up.pdf} in App. A
    (right panel).}
    \label{fig:projections.gluon-parton}
\end{figure}

In order to study the central region around the dead cone in greater detail we consider in Fig.~\ref{fig:projections.gluon-parton} (left channel) the distribution of the primary gluon jets, as obtained from partons, along the $X$ axis in a horizontal stripe of width $-\frac{1}{2}< Y < +\frac{1}{2}$ of the Fig.~\ref{fig:scatter_gluon}, upper row.
The dip of the distribution visible for the stable top-quark at the center at $X_g=0$ is largely filled for the unstable top-quark. 
The maximum reflects the edge of the b-jet where gluons emitted from the b-quark are outside the jet cone of radius R=0.2. These gluons from b-decay are restricted to $X_g<0$ because of angular ordering (see Sect.~\ref{sec:theory}). There is some flow beyond this limit especially for the smallest angles $X_g$  which could come from a mismatch of partons generated by the MC and partons combined into a jet. While for large b-quark jet angles $X_b<-2.5$ (green data points) the dead cone is indicated as dip around $X_g=0$, the evidence is reduced for smaller angles $X_b$. At the smallest angle $X_b<-0.5$ there is some excess of parton flow into the right hemisphere $X>0$. 

\begin{figure}[t]
\includegraphics[height=6cm,width=8.0cm]{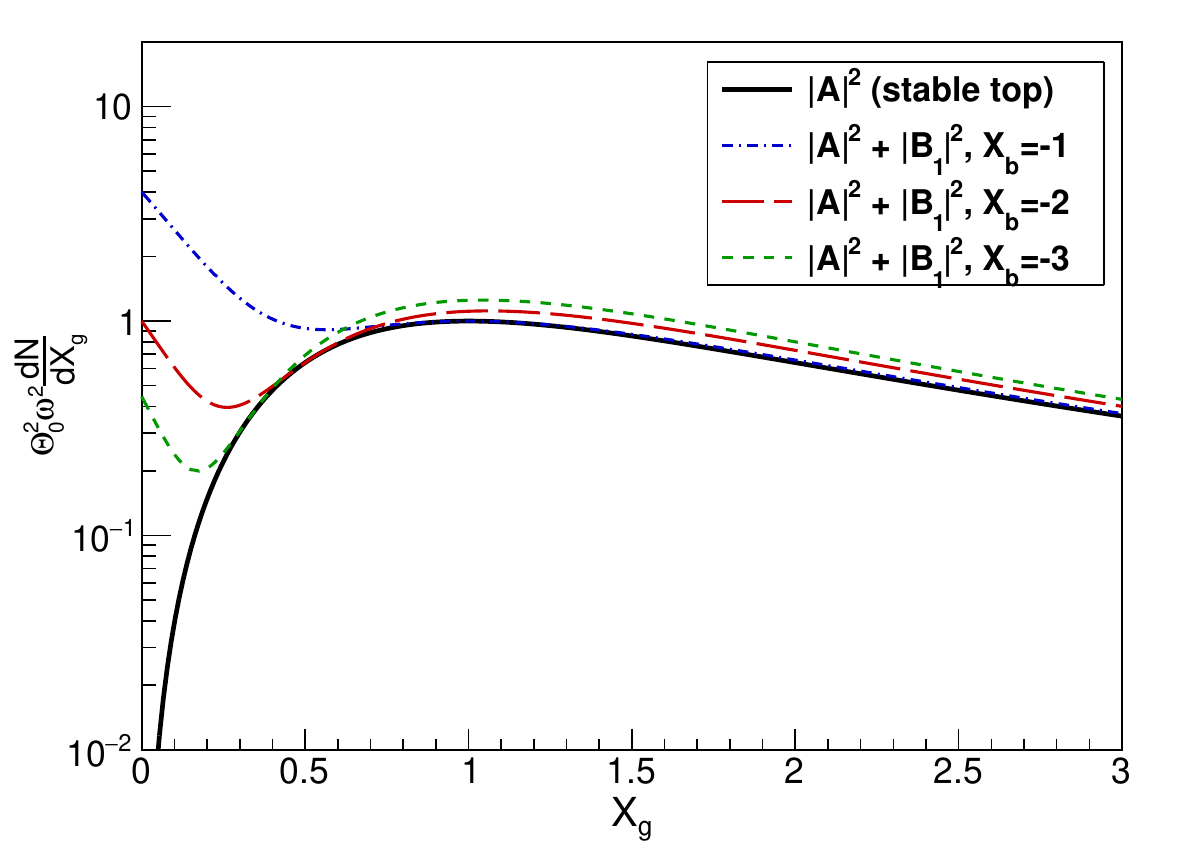} 
  \caption{Distribution of gluon jets along the $X$-axis of the $(X.Y)$ scatterplot according to Eq.~\eqref{eq:X-stripe} for different b-quark angles~$X_b$ with contributions from amplitude $A$ as for stable top-quark with dead cone and from $B_1$ representing the $\widehat{tb}$ dipole. 
} 
\label{fig:X-model}
\end{figure}

The behavior of these distributions can be qualitatively described in terms of the superposition of the two dipoles $\widehat{t \bar t}$ and $\widehat{tb}$ represented by the amplitudes $A$ and $B_1$ in Eq.~\eqref{FKhozeOtt}. The contribution of the b-quark amplitude $B_1$ is not directly known, as we have reduced its contribution by our vertical mass cut, so we calculate this contribution with the full density $|B_1|^2$ for reference. The distributions $dN/dX_g$ along the horizontal stripe can be calculated with $\Theta_g= X_g\Theta_0$ and $Y=0$. Furthermore, we have to express the variables $\Theta',\phi'$ for gluons referring the b-quark direction by the angles $\Theta,\phi$ in reference to the t-quark. Because of the strong collimation of the radiation from the b-quark, we approximate 
\begin{equation}
    \Theta_g'\approx\Theta_g-\Theta_b
\end{equation}
for the region $X_g>0$ opposite to the b-quark direction.
For small angles and high energies with $\beta\approx 1-\frac{1}{2} \Theta_0^2$, $\beta'\approx1$, $\Theta_0'\approx 0$ and noting that $dXdY=d\cos\Theta d\phi\approx \Theta d\Theta d\phi$ we obtain the result for $dN/dX_g\propto F$ in the top-quark hemisphere using Eqs.~\eqref{termA} and~\eqref{termB}
    
\begin{eqnarray}
   \frac{dN}{dX_g}&\propto&  |A|^2+|B_1|^2 \label{eq:X-stripe} \\
    \Theta_0^2\omega^2 |A|^2 &=& \frac{4 X_g^2}{(X_g^2+1)^2},\nonumber\\
    \Theta_0^2\omega^2 |B_1|^2 &=& \frac{4(1+X_g X_b)^2}{(X_g^2+1)^2(X_g-X_b)^2}.
    \nonumber
\end{eqnarray}
and this distribution is presented in Fig.~\ref{fig:X-model} in the region $X_g>0$.

The figure shows that the dead cone, i.e.\ the dip at the center $X_g=0$ of the $A$ amplitude (stable t-quark) is filled by the contribution from the $\widehat{tb}$ dipole (amplitude $B_1\propto 1/X_b^2$ at $X_g=0$ decreasing with increasing $X_b$). On the other hand, at larger angles $X_g$ the contributions from the large b-quark jet angles $X_b$ dominate. 
This increase is illustrated by the respective values of $dN/dX_g$ for the gluon angles $X_g=1$ and $2$ in Tab.~\ref{tab:dNdXg_histogram_values}.

The data in both panels of Fig.~\ref{fig:projections.gluon-parton} show indeed, that at the center $X_{g,p}=0$ the lowest density appears for the largest angle $X_b$ of the b-quark jet (green curve). 
With increasing gluon or parton  angles $X_{g,p}$ the curves cross and the densities $dN/dX_g$ and  $dN/dX_p$ dominate for the largest angle $X_b$.
The effect is more evident for the parton final state in the right panel of Fig.~\ref{fig:projections.gluon-parton}. The same effect is observed for the final state hadrons, see
Fig.~\ref{fig:projections.parton} of App.~\ref{app:angle}. 

This dependence on angle $X_b$ is contrary to the naive expectation that in the sequence of top-quark decay into b-quark both quarks fragment independently, so that for increasing  decay angle the b-fragments move preferentially in the b-quark direction (to the left) and less frequently into the away side (to the right) and that the top-quark  fragments become better visible in the forward direction.
However, the gluon emission from the b-quark does not enter the right hemisphere because of angular ordering and the gluon radiation to the right hemisphere caused by the $\widehat{tb}$ dipole does not originate from the b-quark but from the top-quark and this radiation increases with increasing b-quark angle.

\begin{table}[htbp]
  \centering
  \begin{tabular}{c|c|cccc}
    \hline
    $X_{g}$ 
    & $\displaystyle \frac{dN}{dX_{g}}(X_{g})\big|_{\text{stable top}}$
    & $-1 < X_{b} < -0.5$
    & $-1.5 < X_{b} < -1$
    & $-2.5 < X_{b} < -1.5$
    & $-4 < X_{b} < -2.5$ \\
    \hline
    1.0 & 0.072 & 0.081 & 0.082 & 0.087 & 0.098 \\
    2.0 & 0.051 & 0.056 & 0.057 & 0.058 & 0.059 \\
    \hline
  \end{tabular}
  \caption{
    Values of gluon distribution $\frac{dN}{dX_{g}}$ at $X_{g}=1$ and $X_{g}=2$ for the stable top-quark and for the unstable top-quark in four intervals of angle $X_{b}$ of the b-quark jet, data as in left panel of Fig.~\ref{fig:projections.gluon-parton}.
  }
  \label{tab:dNdXg_histogram_values}
\end{table}

\subsection{Summary on angular distributions}

We summarize our observations on the angular structure of the top-quark jet in the $(X,Y)$ scatter plots as follows. A broad distribution of particles extends over a ring of radius $R_X\sim1$ in rescaled angles $\Theta/\Theta_0$, and a depletion occurs in the center along the direction of the top-quark. This dominant ``dead cone" structure is related to the initial $\widehat{t\bar t}$ dipole. It is visible most clearly for the primary gluon jets emitted from the stable top-quark; because of the contribution from the $\widehat{tb}$ dipole it is less pronounced but still visible  for the primary gluon jets in the unstable top-quark, and it is further reduced for partons and becomes invisible for hadrons. These observations refer to  events selected with lateral b-quark emission and with invariant masses $M(b\ell \nu)$ or $M(B\ell \nu)$ near the top-quark mass $m_t$, i.e., with $t\to tg$ emission enhanced.

Clear evidence is provided for the weaker gluon radiation from the $\widehat{tb}$ dipole as expected from Eq. \eqref{termB}. This is concluded, as with an increasing b-quark jet angle $X_b$ also the primary gluon density  and the parton and hadron densities increase in the region opposite to the b-quark direction, i.e. in the right hemisphere $X>0$. Radiation from the b-quark is limited to the left hemisphere $X_g<0$ due to angular ordering; there is only a small spillover into the right hemisphere limited to an angular region near the center at $\Theta=0$.  Therefore, only the radiation from the top-quark in the $\widehat{tb}$ dipole is present in the right hemisphere as background to the leading dead cone process related to the $\widehat{t\bar t}$ dipole.

In consequence of these findings, we will construct in the following
the inclusive momentum distributions of partons and hadrons for the top-quark fragmentation only from the particles in the right hemisphere with $X_{p,h}>0$ shown  in Fig.~\ref{fig:figure8-10up.pdf} of App. A 
and multiply it by 2; in this region the b-quark radiation is suppressed and there is only the weak radiation from the $\widehat{tb}$ dipole as background to the leading dead cone process looked for, and this background vanishes approximately for zero b-quark jet angle $X_b$. 

\section{Distribution of particle momenta in top-quark jets and MLLA predictions}
\label{sec:momenta}

\subsection{Reconstruction of parton and hadron momentum distributions}
\label{sec:partonhadronmomentumdists}

The study of the angular distribution of particles around the top-quark direction provides evidence for the dead cone effect as modeled by the \textsc{Pythia}~8.3 MCEG. Another approach is the analysis of particle distributions in momentum space that does not require precise knowledge of the top-quark flight direction as in the angular analysis. 
Furthermore, for the momentum spectra there are predictions based on the MLLA which we have explored previously for c- and b-quark jets in $e^+e^-$ annihilation~\cite{Kluth:2023umf}, also in comparison with \textsc{Pythia}~\cite{Kluth:2023dav}. For a stable top-quark the analysis could proceed in the same way and there are well-defined expectations 
for the dependence of the parton momentum distribution on the heavy quark mass in the process $e^+e^-\to t \bar t+X$ related to the primary $\hat{t \bar t}$ color dipole radiation.

The momentum distributions are studied using the variable $\xi=\ln{(1/x)}$ with the fractional momentum or energy~$x$; the MC data are presented as functions of $x_p=2p/W$ at energy $W=\sqrt{s}$  in the c.m.s. or $\xi_p=\ln{(1/x_p)}$. In $e^+e^-$ annihilation, we also refer to the fragmentation function 
\begin{equation}
\bar D(\xi,W)=\frac{1}{2}\frac{1}{\sigma_{tot}}\frac{d\sigma^{e^+e^-}}{d\xi} (\xi,W) 
\label{dbardef}
\end{equation}
with the particle density distribution $d\sigma/d\xi$ in both hemispheres and the total production rate for the given event sample $\sigma_{tot}$ and $\bar D(\xi,W) = x D(x,W)$ with the inclusive distribution $D(x,W)$ for one jet. In this section, we refer to the $\xi$-spectra as $2 \bar D$ for both hemispheres, following the convention in our previous work.

The $\xi_p$-distributions presented here refer to partons or to hadrons with all charges included to be naturally compared. In the experiment, the charged particles are much better reconstructed than the neutral ones (such as the $\pi_0$'s). Therefore, it may be indicated to construct the $\xi_p$-spectra of the hadrons from the charged particles only and to correct for the neutrals. In fact, the ratio $r_{ch/all}$ of the spectra of charged hadrons and all hadrons is rather universal, and it is largely independent of the momentum in the region of our interest. For the top-quark, we find from the MC 
\begin{equation*}
r_{ch/all}=\left\{ \begin{array}{rcl} 0.63 \ \ \ \quad\qquad& \mbox{for} & 2\le\xi_p< 7 \\ 
   0.78 -0.021\ \xi_p & \mbox{for} & 7\leq\xi_p<9\end{array}\right.
\end{equation*}
and the same ratio is also found for the u-quark jet within $0.5\%$ in this region.
We note that MLLA predictions for $\xi_p$-spectra could also be performed directly with charged particles as in previous studies \cite{Acosta:2003XYZ,Kluth:2023umf}. 
The study of angular distributions in the last section has revealed that for angles $|X_b|>0.5$ between the top and the b-quark jets, the circular radiation around the top-quark direction 
is visible. In order to minimize the influence of the radiation from b-quarks, we calculate the momentum spectra of partons or hadrons from top-quark fragmentation only from those with $X_{p,h}>0$ opposite to the b-quark direction and multiply the result by the factor~2.

At first, our aim is to reconstruct the $\xi_p$-distributions for the unstable top-quark after the contribution of the $\widehat{tb}$ dipole radiation is removed from the top-quark fragmentation region. As a test of this procedure we compare the result for partons with the $\xi_p$-distribution obtained for the stable top-quark. This is the distribution which is used in the MLLA test for the relation between spectra for primary massive quarks with the ones for primary light quarks. 

\begin{figure}[t]
    \centering
    \begin{minipage}[c]{0.7\linewidth}
        \centering
        \includegraphics[width=\linewidth]{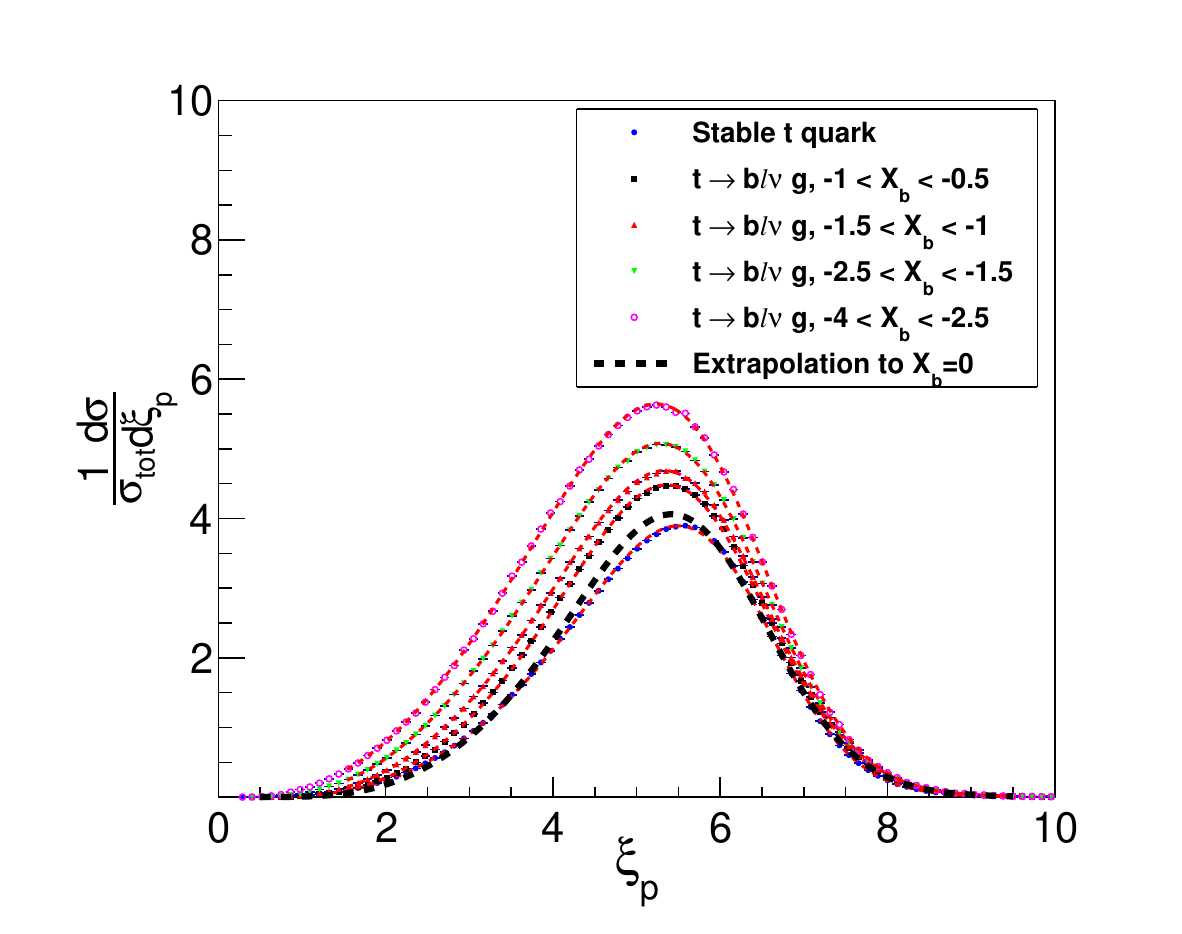}
    \end{minipage}
    \hfill
    \hskip -2.0cm
    \begin{minipage}[c]{0.36\linewidth}
        \centering
        \includegraphics[width=\linewidth]{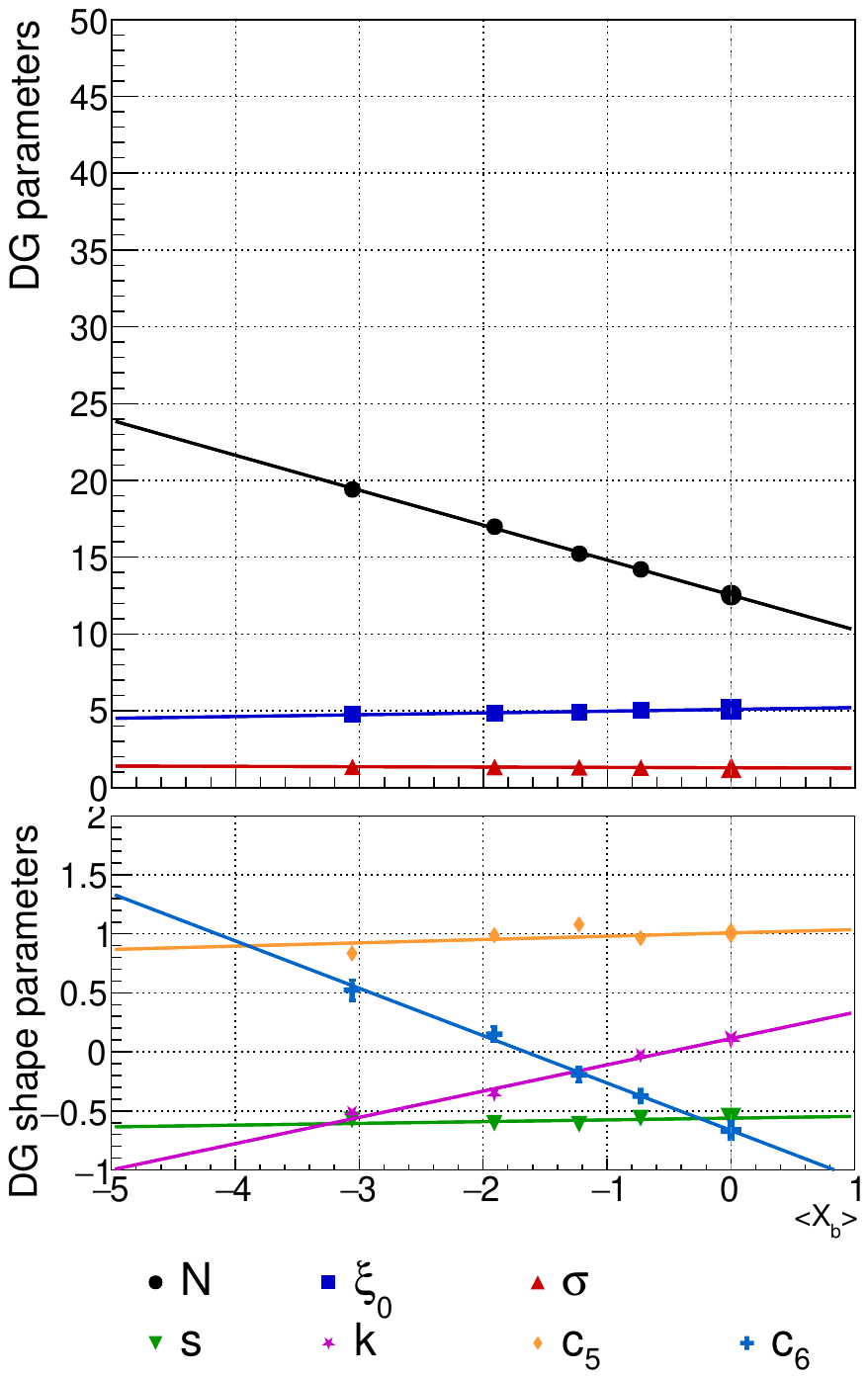}
    \end{minipage}
    \caption{Distribution of partons from top-quark decays over the momentum variable 
    $\xi_p=\ln(1/x_p)$ at c.m.s.\ energy $W=1000$~GeV for different intervals in the decay 
    angle $X_b$ of the b-quark jet from top decay together with fits to the Distorted Gaussian (DG)
    distribution of $O(z^6)$. The dashed curve represents the extrapolation to forward 
    direction $X_b=0$ in comparison with the $\xi_p$-distribution of the stable top-quark 
    (bottom line). The $\xi_p$-distributions are determined from the hemisphere away from the 
    b-quark ($X>0$) and are multiplied by a factor 2; normalization of spectra to $2\bar D(\xi_p,W)$ 
    in Eq.~\eqref{dbardef} (left panel); moment parameters of DG-distribution with fit to linear dependence on b-quark decay angle $X_b$ and extrapolation to the results at $X_b=0$ (right panel).
        }
    \label{fig:multiple_xi_partons}
\end{figure}

\begin{figure}[htbp]
    \hskip -1.1cm
    \begin{minipage}[c]{0.735\linewidth} 
        \centering
        \includegraphics[width=\linewidth]{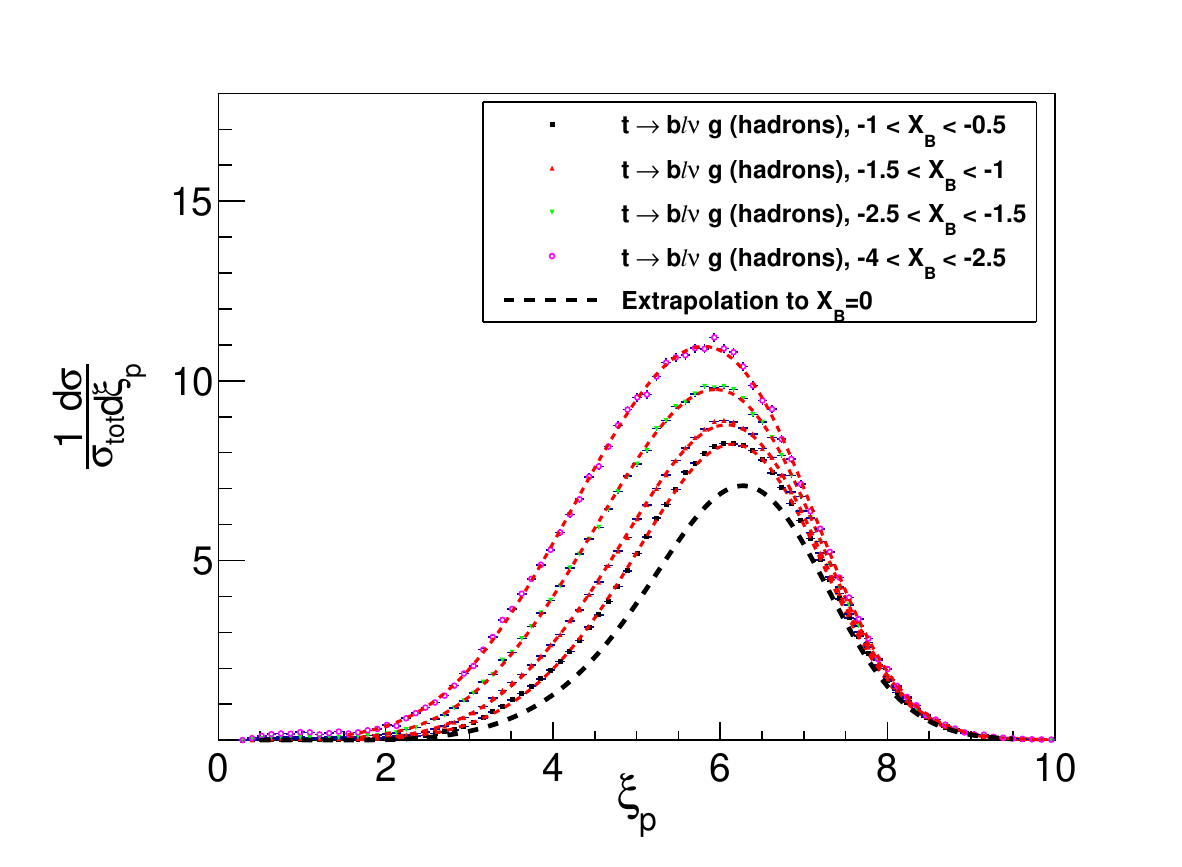}
    \end{minipage}
    \hskip -1cm
   \begin{minipage}[c]{0.30\linewidth} 
       \centering
       \includegraphics[width=\linewidth]{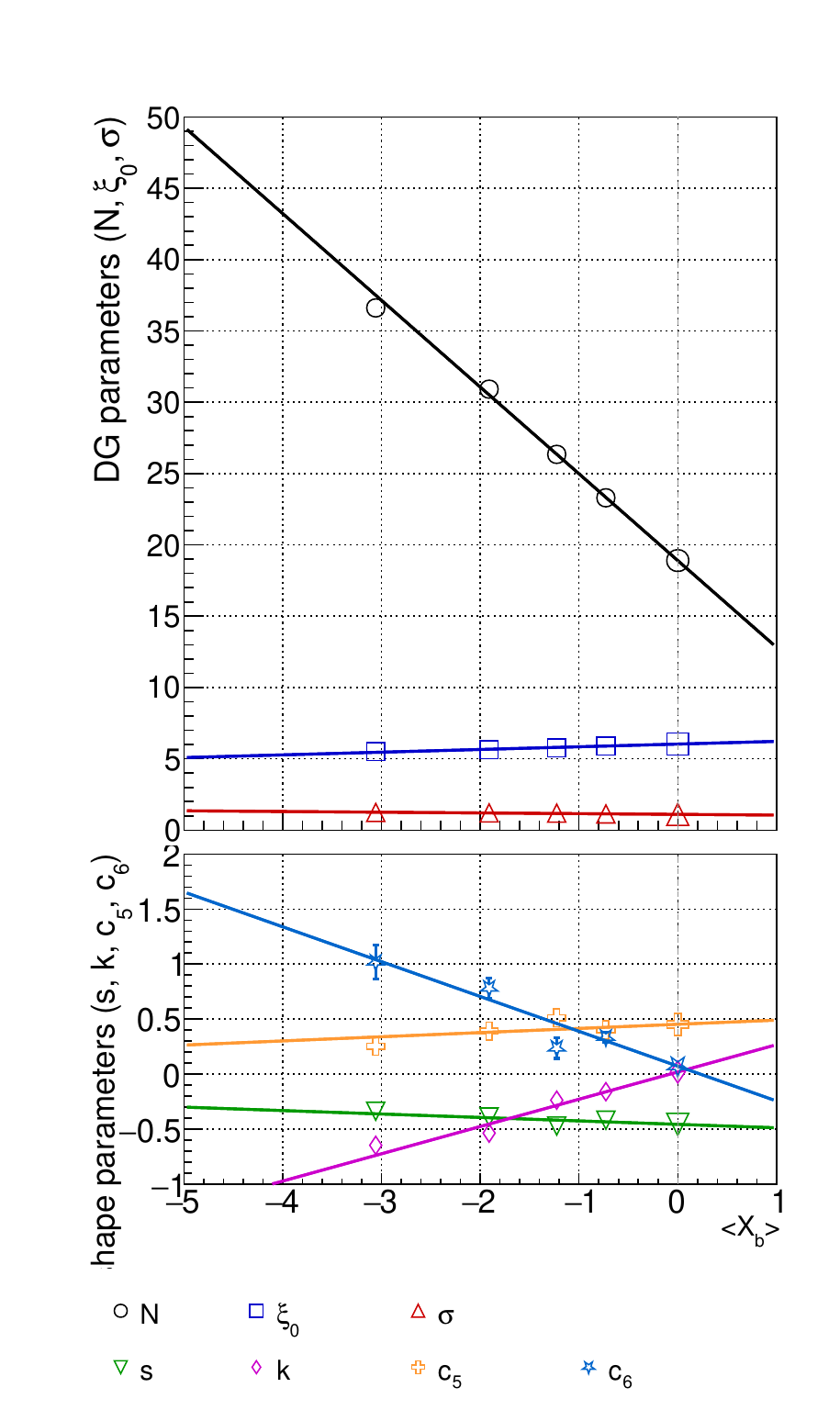}
    \end{minipage}
    \caption{Distribution over $\xi_p=\ln(1/x_p)$ as in Fig.~\ref{fig:multiple_xi_partons} but for hadrons. The $\xi_p$-distributions are shown for different intervals of the decay angle $X_B$ of the b-quark jet from top decay, the dashed line represents their extrapolation towards the  $\xi_p$-distribution at $X_b=0$ with the $\widehat{tb}$ dipole radiation removed; right panel: moment parameters of DG distribution and linear fits to $X_B$ dependence with extrapolation to $X_B=0$.
   }
    \label{fig:multiple_xi_hadrons}
\end{figure}

To this end, we determine the $\xi_p$-distributions for different intervals of decay angle $X_b$ of the b-quark jet and extrapolate these distributions to the top-quark direction $X_b=0$. In this limit the radiation from the $\widehat{tb}$ dipole is expected to vanish at high energies as discussed in Sect.~\ref{sec:theory}, Eq.~\eqref{B1average}. The $\xi_p$-distributions of partons for different angular intervals of $X_b$ are shown in Fig.~\ref{fig:multiple_xi_partons}. The maxima of these distributions decrease with decreasing $|X_b|$. 

The extrapolation of these $\xi_p$-distributions towards $X_b\to 0$ is performed by fitting the distributions to a Distorted Gaussian expansion, and to extrapolate the moment parameters. This representation is known to be well suited for $\xi$-distributions~\cite{Fong:1990nt}. The well known representation up to $4^{th}$ order in $z=(\xi-\xi_0)/\sigma$ fits the $\xi_p$-distributions for $\xi_p\gtrsim3$ only, therefore we added two more terms to fit the full $\xi_p$-region considered, see App.~\ref{app:DG}. The moments are shown in the right panel of Fig.~\ref{fig:multiple_xi_partons}, and the data are well fitted to a linear dependence on $X_b$. From the extrapolated values at $X_b=0$ the corresponding $\xi_p$-distribution is reconstructed and is also shown in Fig.~\ref{fig:multiple_xi_partons} as the dashed curve.
 
The same procedure is also applied to the $\xi_p$-spectra of hadrons, which are selected from the right side hemisphere $X_h>0$ of the scatter plots in the lower panels of Fig.~\ref{fig:figure8-10up.pdf} in App.~A. In Fig.~\ref{fig:multiple_xi_hadrons} we show the results of the fits to the $\xi_p$-distributions by the distorted Gaussian representation. The corresponding seven Gaussian moment parameters are again well fitted by a linear function, 
and in this way the extrapolated moments at $X_B=0$ and the extrapolated $\xi_p$-distribution can be obtained as in the case of partons. This result from extrapolation represents the $\xi_p$-distribution for hadrons from top-quark decays with contributions from the $\widehat{tb}$ dipole subtracted and is shown in Fig.~\ref{fig:multiple_xi_hadrons} as a dashed line.  
\begin{table}[bht]
  \centering
  \small
  \setlength{\tabcolsep}{4pt}
  \begin{tabular}{c cc c c c c c c c c c}
    \hline
    Bin &
    $\langle X_b \rangle_{\text{part}}$ &
    $\langle X_B \rangle_{\text{had}}$ &
    Level &
    $N$ &
    $\xi_{0}$ &
    $\sigma$ &
    $s$ &
    $k$ &
    $c_{5}$ &
    $c_{6}$ &
    $\chi^{2}/\mathrm{ndf}$ \\
    \hline
    1 & -0.728 & -0.728 & partons &
    14.2 & 5.01 & 1.30 & -0.561 & -0.025 & 0.964 & -0.374 & 5.83 \\
     &  &  & hadrons &
    23.3 & 5.90 & 1.13 & -0.420 & -0.161 & 0.401 & 0.333 & 10.1 \\
    \hline
    2 & -1.22 & -1.22 & partons &
    15.2 & 4.92 & 1.32 & -0.612 & -0.196 & 1.07 & -0.194 & 3.65 \\
     &  &  & hadrons &
    26.3 & 5.77 & 1.19 & -0.473 & -0.237 & 0.512 & 0.235 & 3.39 \\
    \hline
    3 & -1.90 & -1.91 & partons &
    16.9 & 4.84 & 1.34 & -0.604 & -0.354 & 0.988 & 0.149 & 2.08 \\
     & &  & hadrons &
    30.9 & 5.64 & 1.20 & -0.390 & -0.537 & 0.388 & 0.781 & 1.95 \\
    \hline
    4 & -3.05 & -3.05 & partons &
    19.4 & 4.77 & 1.35 & -0.575 & -0.512 & 0.834 & 0.522 & 1.78 \\
     & &  & hadrons &
    36.6 & 5.51 & 1.24 & -0.339 & -0.648 & 0.253 & 1.01 & 1.24 \\
    \hline
  \end{tabular}
  \caption{Fit parameters of the Distorted Gaussian distribution of $O(z^6)$ for the $\xi_p$-distributions of unstable top decays at parton and hadron level
  in four bins of the b-quark jet angles $\langle X_{b,B} \rangle$; $\chi^{2}/\mathrm{ndf}$ indicates fit quality.}
  \label{tab:DGz6_moments_parton_vs_hadron_noerr}
\end{table}
\begin{figure}[b!]
   \centering
    \begin{tabular}{cc}
   \hskip -1.1cm 
   \includegraphics[width=0.5\linewidth]{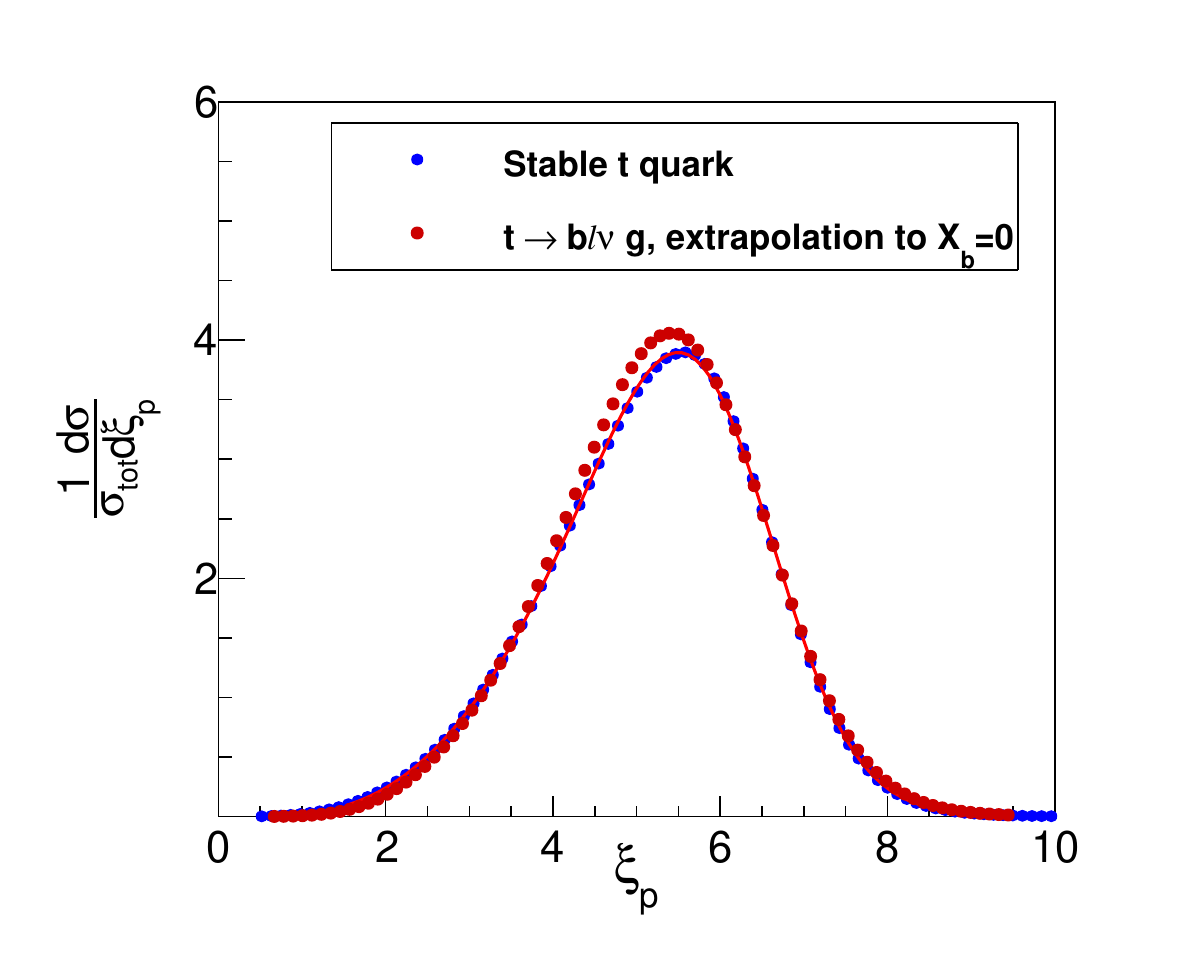}   & 
   \includegraphics[width=0.5\linewidth]{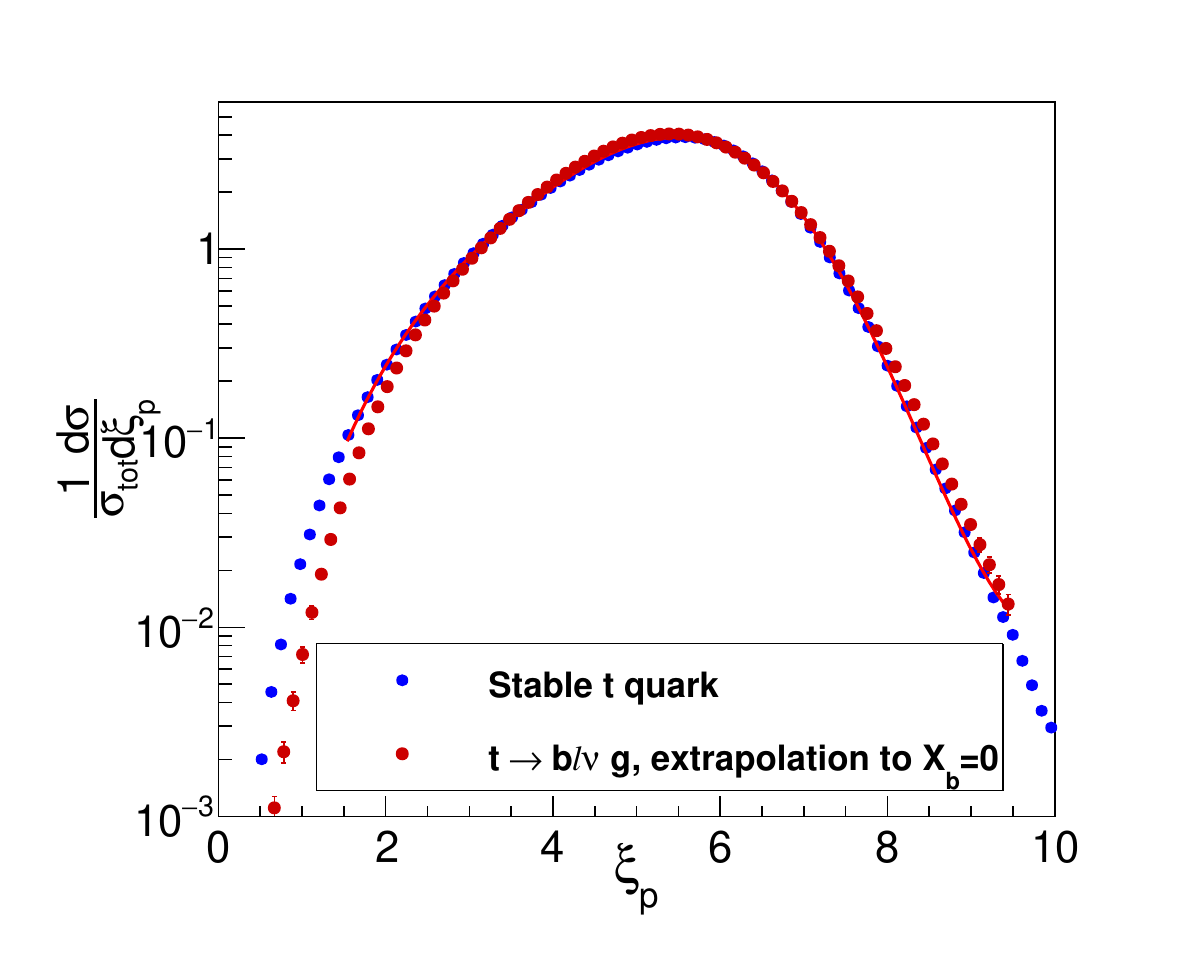}
     \end{tabular}
    \caption{Distribution over $\xi_p=\ln(1/x_p)$ 
    of partons as in Fig.~\ref{fig:multiple_xi_partons} after extrapolation to forward decay angle $X_b=0$ of b-quark jet to remove the $\widehat{tb}$ dipole radiation, in comparison with $\xi$-distribution for stable top-quark as consistency check, based on fits to Distorted Gaussian.
    }
    \label{fig:stable-unstable}
\end{figure}

As a function of the angle $X_b$ of the b-quark jet, the main variation of the $\xi_p$-distributions in Figs.~\ref{fig:multiple_xi_partons} and~\ref{fig:multiple_xi_hadrons} comes from normalization, while the shape varies only slightly. This is reflected in the behaviour of the moment parameters, 
where the strongest variation occurs in the multiplicity parameter $N$ whereas the variation in the second and third moments is relatively small and the higher four moments are numerically small. This behaviour reflects the property of the distributions in leading order $\alpha_s$ to factorize in energy $\omega$ and angle $\Theta_g$, see Eq.~\eqref{FKhozeOtt}, so the shape of the $\xi_p$-distribution in this approximation is independent of the angle $X_b$. 

For numerical values of the moment parameters, see Tab.~\ref{tab:DGz6_moments_parton_vs_hadron_noerr}. There is a larger $\chi^2$-value for the lowest values of the b-quark jet angle $X_b$ which may be related to the influence of the b-quark radiation in the right hemisphere $X>0$ and close to the top-quark direction. An improvement of these fits could be obtained by including even higher moments in the fit; the effect on the lower leading moments is expected to be small. We note that the moments closest to $X_b=0$ (bin 1) fit well with the linear functions in Figs. \ref{fig:multiple_xi_partons} and \ref{fig:multiple_xi_hadrons}, so there is no evidence of an anomaly near the hemisphere boundary. In order to test the validity of our extrapolation procedure, we  compare with the results from the stable top-quark.

In Fig.~\ref{fig:multiple_xi_partons} the extrapolated $\xi_p$-distribution for partons, the dashed line, is also compared with the control curve from the stable top-quark (bottom line). The same comparison is also shown in Fig.~\ref{fig:stable-unstable} in more detail. This is the main test for the success of our extrapolation method. The extrapolation removes the contribution from the $\widehat{tb}$ dipole (the $|B_1|^2$ contribution in Eq.~\eqref{FKhozeOtt}) and the $|A|^2$ contribution for the stable top-quark remains. 
There is a good agreement within about 5\% near the maximum and below 12\% for $\xi_p\gtrsim 3$ within the variation of about 50\% within the full extrapolation range. This indicates that the contributions from the $\widehat{tb}$ dipole are effectively removed, and we consider the remaining difference as a systematic error of the method.

A comparison of the hadron spectra with the stable top-quark as a check is not possible, because there are no hadronic final states with the stable top hadron to be removed. We therefore rely on the successful extrapolation procedure for partons. A comparison between the moments for partons and hadrons shows that the main difference is in the larger multiplicity in an event for hadrons which is expected from hadronization. The behaviour and size of the other parameters are quite similar. A remarkable feature is the linearity in the dependences on angle $X_b$. This dependence appears at the level of parton evolution as is indicated from the behaviour of parton and hadron spectra for $X>0$ in the right panels of Figs.~\ref{fig:projections.gluon-parton} and~\ref{fig:projections.parton}, while the dependence of the gluon spectrum on this angle is rather weak and it is expected to depend quadratically on the decay angle for small angles according to Eq.~\eqref{termB}. 
%

\subsection{Expectations for momentum spectra within the Modified Leading Log Approximation of QCD}

The dead cone effect in heavy quark jets has first been established for the total multiplicities of light particles within the MLLA of perturbative QCD~\cite{Dokshitzer:1991fc,Dokshitzer:1991fd,Schumm:1992xt,Dokshitzer:2005ri}. A corresponding relation for particle momentum spectra is not yet available in MLLA, but an equation for inclusive $x$-spectra has been suggested, which reproduces the equation for total multiplicities after integration over $x$ and avoids a negative fragmentation function. This MLLA expectation has been reported in~\cite{Dokshitzer:1991fd} and can be written in the $\xi$ variable as 
\begin{equation}
    \bar D_Q(\xi,W) = \bar D_q(\xi,W) - \bar D_q(\xi - \xi_Q,\sqrt{e} m_Q)
    \label{DMLLA}
\end{equation}
with $\xi_Q=\ln(1/\langle x_Q\rangle)$ for the heavy quark $Q$. This equation relates the momentum distribution in the heavy quark jet at energy $W$ to the one in the light quark $q$ jet at the same energy after subtraction of the distribution at the reduced scale $W_0=\sqrt{e} m_Q$. This equation has been found to be rather well supported by experimental data on c- and b-quark jets in the central $\xi_p$ region for momenta $p \gtrsim \Lambda_{QCD}$ \cite{Kluth:2023umf}. For the top-quark we derive the numerical value for $\xi_t$ from the top-quark energy distribution obtained using the \textsc{Pythia} 8.3 MCEG and find a rather small shift $\xi_t$
\begin{equation}
    \langle x_t\rangle=0.92,\ \xi_t= 0.083 \ \ \rm{at}\ W=1000\  \rm{GeV}.
    \label{toppar}
\end{equation}
 
\begin{figure}[t!]
\includegraphics[height=10.5cm,width=12.5cm]{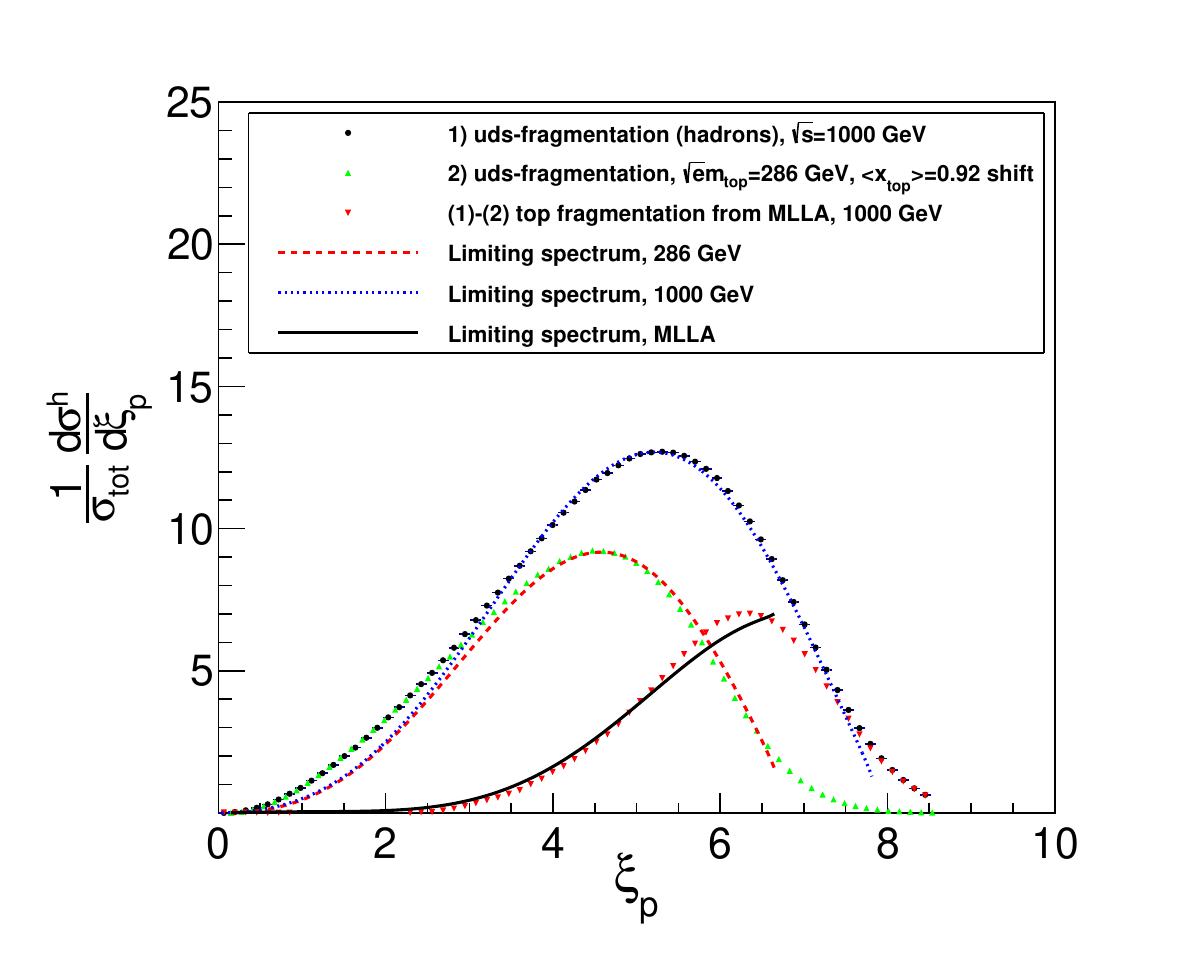} 
  \caption{Prediction for $\xi$-distributions in top-quark fragmentation from MLLA of QCD for the hadrons shown as red triangle according to Eq.~\eqref{DMLLA} using as input the $\xi$-distributions of light uds-quarks at energies $W=1000$ GeV and $W_0=286$ GeV of the \textsc{Pythia} 8.3 MCEG. Also shown are the results from the ``limiting spectrum" of MLLA as curves. 
}
\label{fig:xi-dist stable MLLA 1000}
\end{figure}

Currently, there are no experimental data available on these fragmentation functions to test Eq.~\eqref{DMLLA}. In the following, we investigate how such a test could be performed using results obtained by \textsc{Pythia}~8.3. 
In the application of Eq.~\eqref{DMLLA} we generate the distributions on the r.h.s.\ of Eq.~\eqref{DMLLA} and derive the prediction for the heavy quark distribution on the l.h.s.\ which thereafter we will compare with the results obtained in the last subsection on the unstable top-quark. The elements of this prediction are shown in Fig.~\ref{fig:xi-dist stable MLLA 1000}. The subtraction of the two distributions for light quarks at scales $W$ and $W_0$ yields the prediction (red triangles) with strong suppression at small $\xi_p$.

We also show the results on the analytical calculations of the $\xi$-spectra in the MLLA of QCD, called ``Limiting Spectrum"~\cite{Dokshitzer:1992jv,Dokshitzer:1992jv}, for the two energies $W$ and $W_0$. These functions are defined in terms of two parameters, the QCD scale $\Lambda_{QCD}$ and normalization $K$. In the derivation of the ``Limiting Spectrum" it is assumed that the partonic branching process proceeds down to a hadronic cutoff $Q_0=\Lambda_{QCD}$ and this result can be compared directly to the spectrum of hadrons in accordance with the Local Parton Hadron Duality (LPHD) hypothesis~\cite{Azimov:1984np,Azimov:1985by}. In order to obtain a good fit to the \textsc{Pythia}~8.3 data, we adjusted the parameters slightly in comparison to our previous applications in~\cite{Kluth:2023umf}, see Table~\ref{LS_Table}. This small variation of parameters reflects higher order perturbative contributions necessary for this large energy interval which are included in \textsc{Pythia}.

\begin{table}[htbp]
\centering
\large
\caption{${\cal K}$ and \(\Lambda_{\mathrm{QCD}}\) values for the relevant energy scales.}
\label{tab:Kch_values}
\rowcolors{2}{gray!15}{white}
\begin{tabular}{|c|c|c|}
\hline
\rowcolor{gray!40}
\textbf{\(\sqrt{s}\) (GeV)} & \textbf{\(\Lambda_{\mathrm{QCD}}\) (MeV)} & \textbf{\({\cal K}\)} \\
\hline
91.2  & 175 & 1.33  \\
286   & 175 & 1.15  \\
1000  & 200 & 1.105 \\
\hline
\end{tabular}
\label{LS_Table}
\end{table}
A remarkable feature is the prediction of a strong suppression of particle production in the heavy quark jet at small $\xi$ as compared to the light quark jet which reflects the dead cone effect for the heavy mass. On the other hand, at large $\xi$ both distributions approach each other, as the soft particles are emitted in a universal way independent of the quark mass.
\subsection{Test of MLLA predictions for top-quark fragmentation}

\begin{figure}[htbp!]
\includegraphics[height=8.8cm,width=9.0cm]{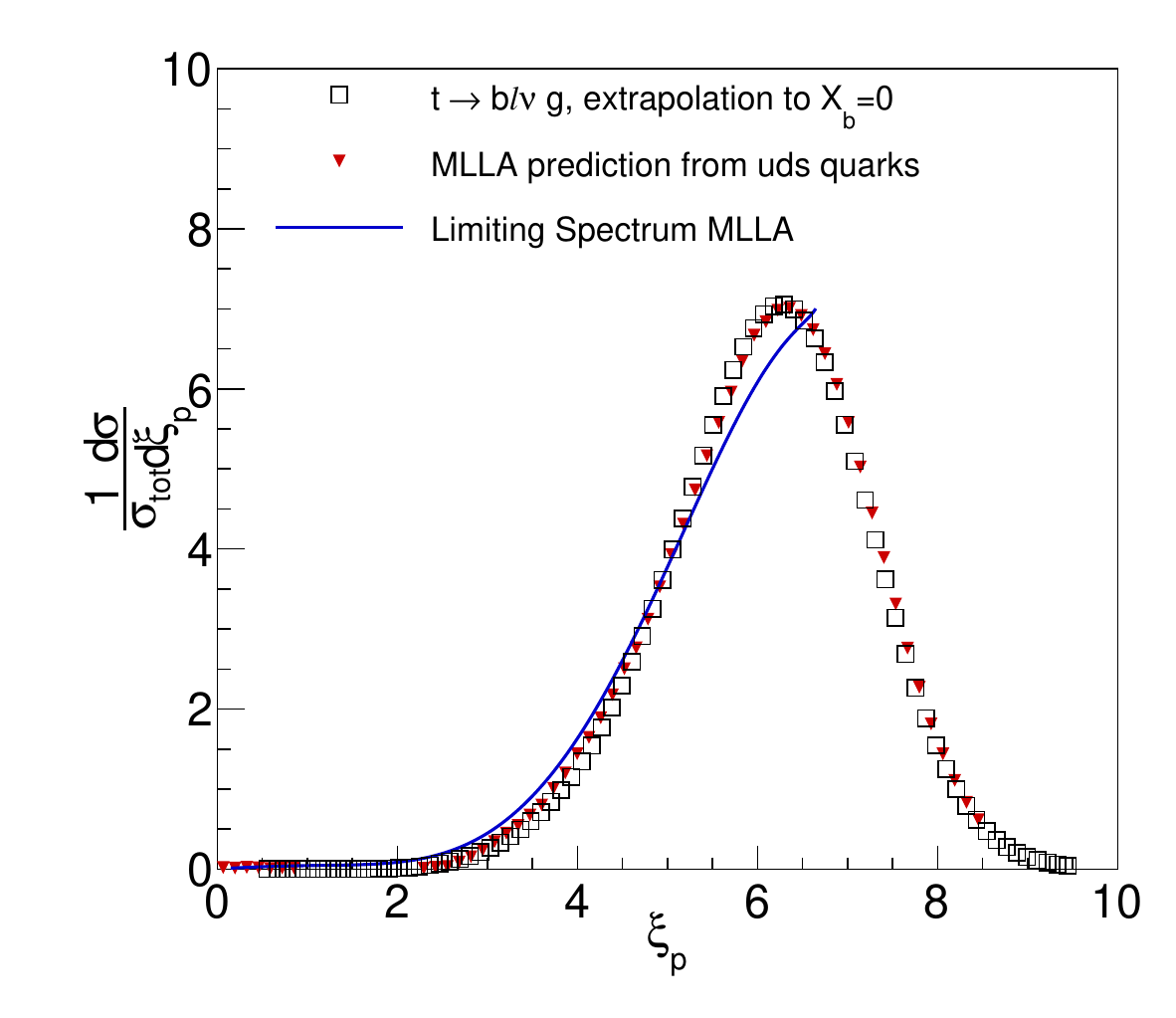}
\hskip - 0.3cm
\includegraphics[height=8.8cm,width=9.0cm]{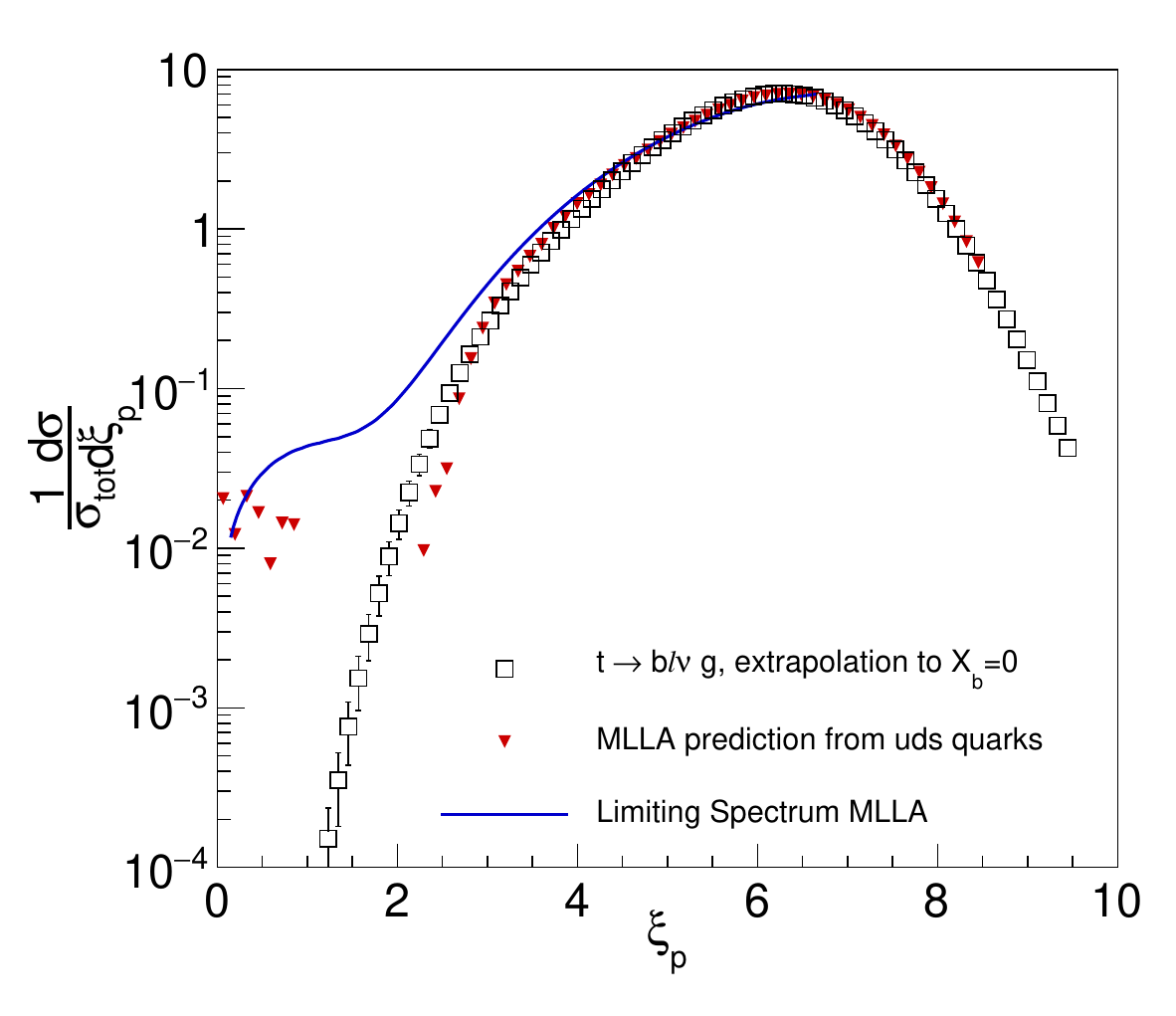}
  \caption{$\xi_p$-distributions of hadrons for top-quark fragmentation obtained by extrapolation to $X_b=0$ in comparison with  the MLLA expectation from $\xi_p$-distributions from light quarks according to Eq.~\eqref{DMLLA}. The curves obtained from the ``Limiting Spectrum" are also shown on both panels.
}
\label{fig:xi-dist-MLLA 1000}
\end{figure}

Finally we compare the results obtained from the angular extrapolation on the top-quark fragmentation into hadrons in subsection~\ref{sec:partonhadronmomentumdists} with the MLLA predictions. This is shown in Fig.~\ref{fig:xi-dist-MLLA 1000}.
The reconstructed distribution comes close to the distribution expected from the MLLA prediction (red triangles in Fig.~\ref{fig:xi-dist-MLLA 1000}) in a large region for $\xi_p>3$ to within about 15\%. A failure of the MLLA approximation at very small $\xi_p$-values (high momenta) should not be surprising, as this approximation is developed for the bulk of particles with lower energies.

\begin{figure}[t!hbp]
\includegraphics[height=7.5cm,width=8.7cm]{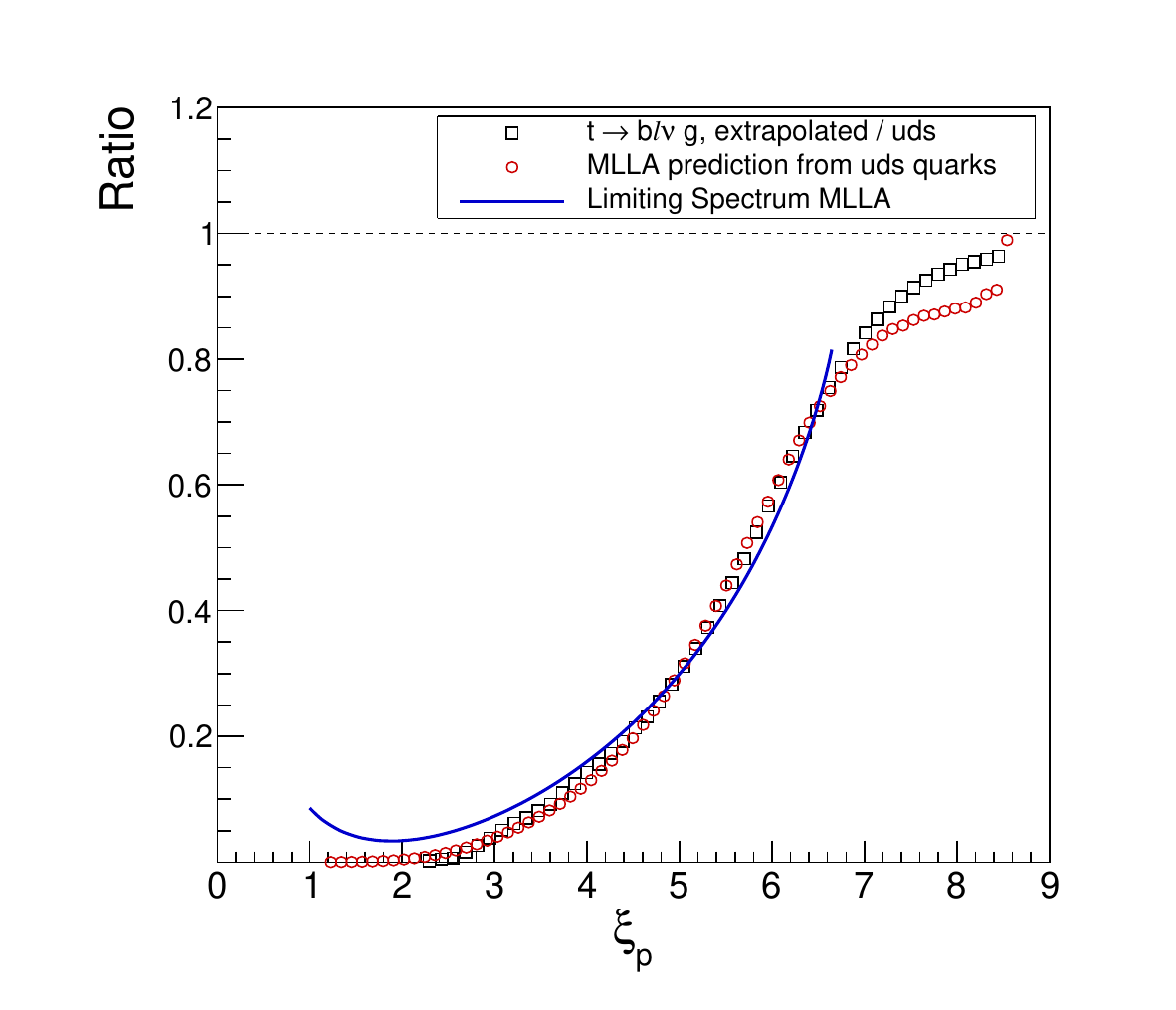}
\includegraphics[height=7.5cm,width=8.7cm]{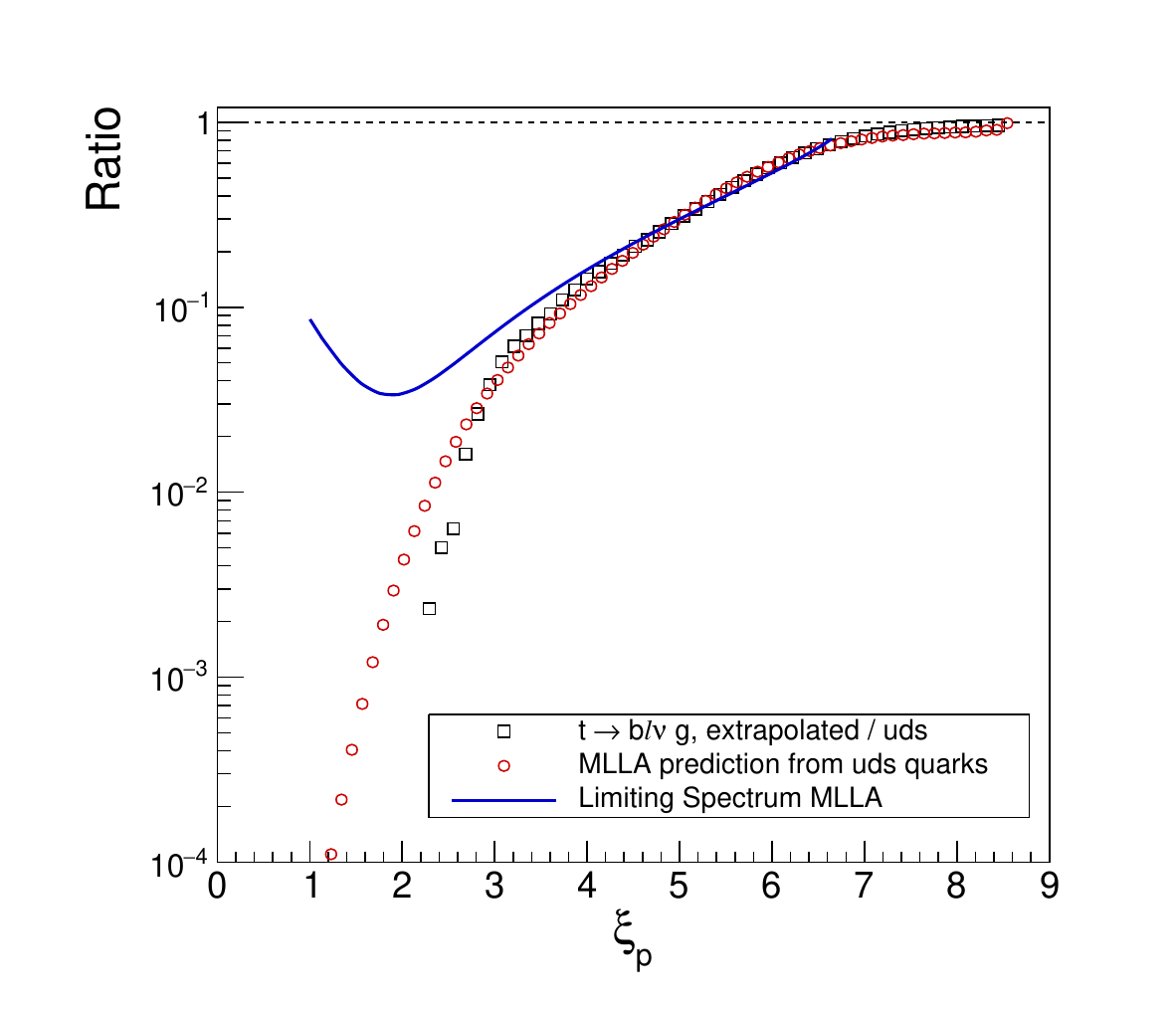}
  \caption{Ratio of the top-quark fragmentation functions as in Fig.~\ref{fig:xi-dist-MLLA 1000} over the fragmentation function for the primary light uds quarks showing the strong suppression of top-quark fragmentation at small $\xi_p$ or large momenta by a factor of about 100, also shown is the  MLLA prediction using Limiting spectrum functions (full curve).
}
\label{fig:xi-dist-MLLA 1000-h1}
\end{figure}

The main result of the study concerns the strong suppression of energetic particles with high energy and its understanding within QCD in terms of a simple connection with spectra from primary light quarks near the energies of the collision $W$ and of the heavy quark mass $m_Q$. The ratio of the expected top-quark fragmentation function to the light quark fragmentation function at $W=1000$~GeV for hadrons is shown in Fig.~\ref{fig:xi-dist-MLLA 1000-h1}. It approaches 1 at large $\xi_p$ for soft particles and decreases down for fast particles at $\xi_p\sim 2$ by a factor of $\sim 100$, again with a good agreement between the top-quark fragmentation and MLLA prediction from light quark fragmentation.

In an alternative approach to establish the dead cone effect, angular spectra instead of momentum spectra have been studied. The QCD prediction in Eq.  \eqref{emission} refers to the angle $\Theta$ between the gluon and the top-quark direction, but the top-quark direction is not well determined by experiment. Therefore, its fluctuation is simulated by MC, thereby the observable forward depletion expected from Eq. \eqref{emission} is reduced.  In case of the top-quark the radiation from the $\widehat{tb}$ dipole contributes to the background which is maximal at small angles (see for example Fig. \ref{fig:projections.gluon-parton}). 
It is an advantage of the study of the momentum spectra that this weakening of the dead cone effect can be reduced and the sensitivity to the quark mass effect be increased.
For c- and b-quark jets in $e^+e^-$ annihilations we found a suppression of high momenta by down to a factor 1/10 \cite{Kluth:2023umf} and in the present MC study a suppression for top-quarks of down to 1/100 (see Fig. \ref{fig:xi-dist-MLLA 1000-h1}). 
The production of gluon subjets at small angles for primary c-quarks reported from the ALICE experiment at LHC \cite{ALICE:2021aqk} showed a reduction of approximately  1/2. For the top-quark, a similar suppression has been found in the MC study  
for $pp$-collisions at 13 TeV \cite{MaltoniSelvaggiThaler:2016DeadCone}. 
The procedures based on angular and momentum studies both predict the observable effects of the dead cone for the top-quark production; the sensitivity is estimated from the MC but it will depend also on the experimental conditions.

\section{Modifications of particle spectra due to finite width of top-quark}
\label{sec:finite_width}

In our discussion so far we have assumed that production and decay of the top-quark are separated by a large time interval and therefore there is no interference between the radiation emitted before and after decay. This corresponds to the default version of the \textsc{Pythia} 8.3 MCEG. Recently, finite width effects in resonance decays have been included in parton showers in an ``interleaved" treatment~\cite{Brooks:2021kji} based on the Vincia shower framework~\cite{Brooks:2019xso}. In this approach the parton branching process at low intrinsic $p_T\lesssim \Gamma_t$ is modified and a suppression of final state particles with energies around the width $\Gamma_t$ by a small amount is obtained. We investigate the consequences of this approach which is available in \textsc{Pythia} 8.3 with flag TimeShower:recoilStrategyRF~\cite{SkandsPrivate}.

\begin{figure}[htbp]
    \centering
    \includegraphics[width=0.45\linewidth]{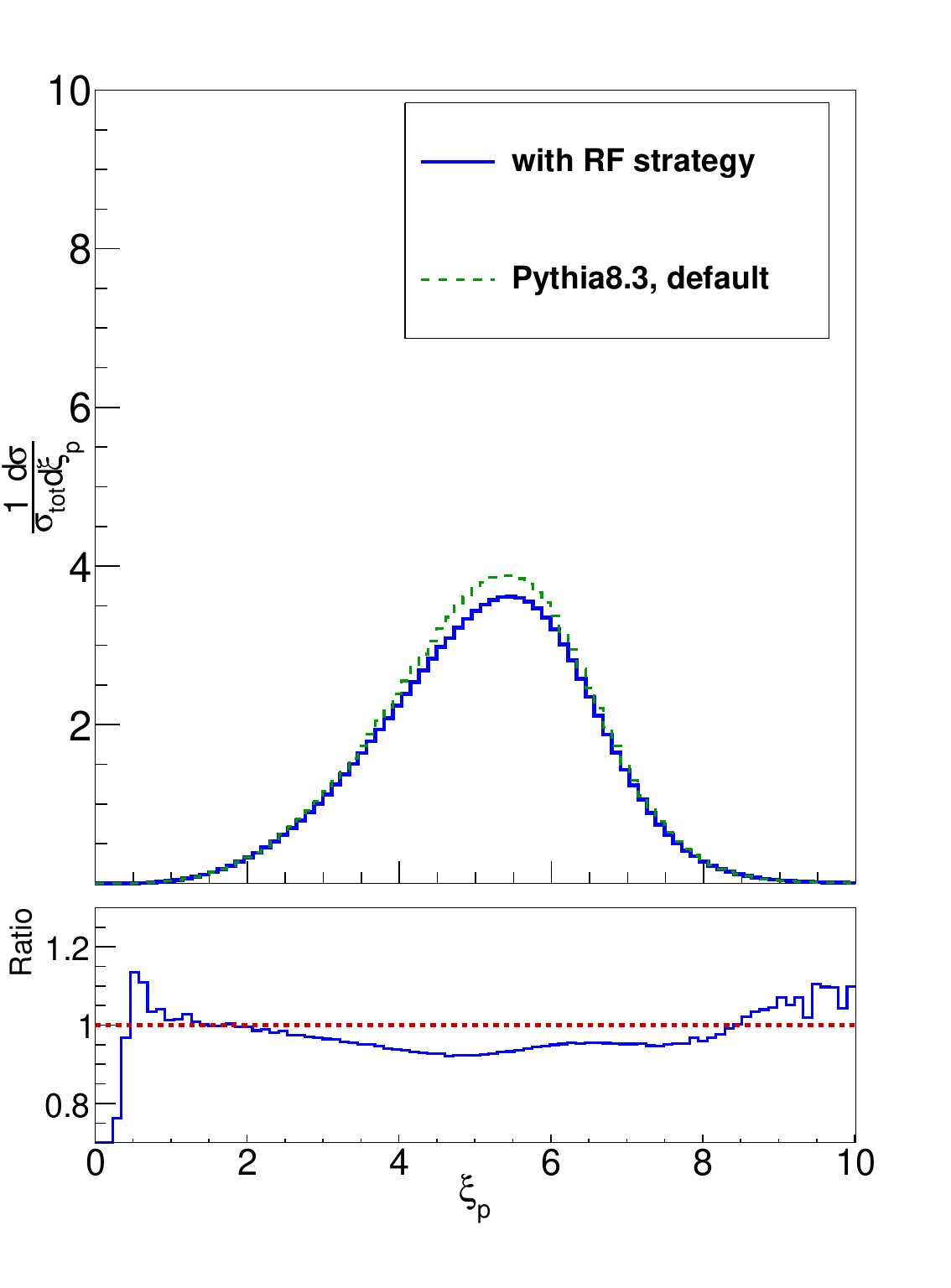}
    \includegraphics[width=0.45\linewidth]{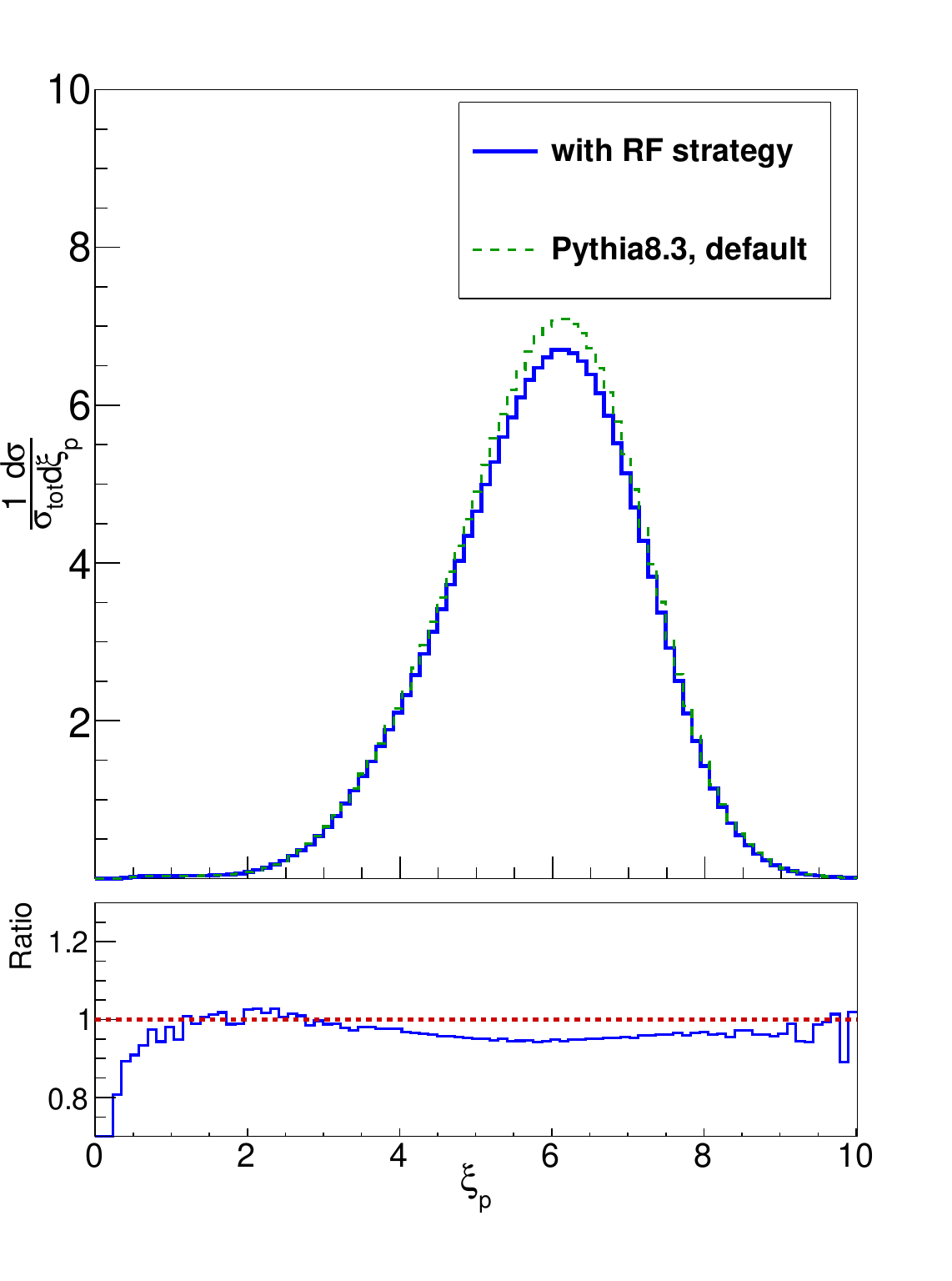}
    \caption{$\xi_p$-distribution of partons (left panel) and hadrons (right panel) with and without finite width correction according to the RF strategy of \textsc{Pythia} 8.3. The maximal effect occurs near $\xi_p=\ln(E/\Gamma_t) \sim 5.8$ at the jet energy of $E=500$ GeV.}
    \label{fig:xi-spectrum with-RS}
\end{figure}

In this way we obtain the modification of the $\xi$-spectra due to the finite top-quark width shown in Fig.~\ref{fig:xi-spectrum with-RS}. This modification results in a small reduction in the rate of $\lesssim 10\% $ at $\xi\sim 5.8$, the value corresponding to $\xi\sim\ln{(500\ \rm{GeV}/\Gamma_t) }$, while there is no change below $\xi\sim 3$. There is an indication of a small increase at very large $\xi$, as also obtained in~\cite{Dittmaier:2025htf}. The data show the result for the inclusive $\xi_p$-distribution, additional cuts did not produce major significant changes. The predicted effect is below the systematic errors that come from the subtraction of $\widehat{tb}$-dipole radiation and is therefore not detectable within our analysis of the momentum spectra.  

\section{Applications to $pp$ collisions - an outlook}
\label{sec:pp}

So far we have discussed top-quark production in the reaction $e^+e^-\to t \bar t + X$, which is most easily formulated theoretically and explored in experimental analysis. There we considered the hadronic final state in the hemisphere of the produced top-quark. Unfortunately, the related measurements will not become available for some time, if ever. Therefore, in this section, we will discuss some strategies to verify our results by measurements feasible at the Large Hadron Collider. The ATLAS and CMS collaborations have published data on $t\bar t$ production in the LHC obtained with triggers on $e^\pm,\mu^\pm$ from $W$-decay or on purely hadronic jets. Cross sections have been measured for the 
transverse momentum of the top-quark up to $p_T\sim 1000$ GeV and the rapidity within $\lvert y \rvert<2.5$, also the top mass $m_t$ and $\alpha_s$ have been determined, for results obtained at 13~TeV, see e.g.\ Refs.~\cite{ATLAS:2017eqx,CMS:2017DiffTTbarLJ,CMS:2019XsecMassAlphaS,ATLAS:2020aln}.

In applying our results to $pp$-collisions, we have to take into account that, due to the restricted rapidity range for jet production, the half opening angle $\Theta_{jet}$ of the top-quark jet has to be smaller than $\pi/2$ as for the full top-quark hemisphere. Furthermore, the background of particles inside the jet cone from underlying event interactions and from initial state radiation
has to be taken into account. 
 Pile-up mitigation at the LHC is effective for charged particles by associating them to the primary vertex of the hard interaction. Pile-up collisions have primary vertices separated along the beam direction and this can be measured well by the semiconductor based
vertex detectors of the experiments; the contribution of photons from pile up events can be reduced in our event sample through the mass cut on $M(b\ell\nu)$ around the top-quark mass. 
An event sample convenient for further analysis would be a final state $p p\to t \bar t + X$ with one top-quark decaying hadronically and the other through $t\to b \ell \nu$. We consider here two possible strategies for the analysis following previous investigations.

The first strategy aims at a full correction of the top-quark observables considered for the ``backgrounds" above. In a study of high $p_T$-jet production, the CDF-collaboration at the Fermilab Tevatron $p\bar p$ collider~\cite{Acosta:2003XYZ} has determined momentum distributions in the $\xi$-variable at different jet energies and jet opening angles. The background  in the jet from the above sources and others 
has been determined experimentally from a control angular region. After subtracting this background from the raw $\xi$-distribution, a corrected distribution has been obtained and is successfully compared to the theoretical predictions from MLLA. The correction was necessary for low momentum particles in a $\xi$-interval of length $\Delta \xi\sim 2$ below  the maximum $\xi$-value at a maximal height of $\sim 10\%$ of the maximum of the $\xi-$distribution. Alternatively, the contribution of 
background to the $\xi$-distribution could be obtained from  a MC simulation.

In an application of this strategy to the study of the top-quark dead cone, the opening angle $\Theta_{jet}$ should be chosen larger than the dead cone angle $\Theta_0=m_t/E_t$, an angular size of $\lvert X \rvert,\lvert Y \rvert<R_X$ with $R_X=2-3$ appears appropriate from Fig.~\ref{fig:ring}, which corresponds to a jet opening angle $\Theta_{jet}=(2-3)\ \Theta_0\approx 40^\circ-60^\circ$ at a jet energy of $E_t=500$~GeV as we had chosen above for $e^+e^-$ annihilation. At the LHC the production of ``fat jets" with jet radius $R=R_X \Theta_0>1$ corresponding to $\Theta_{jet}>57^\circ$ has been discussed e.g.\ in~\cite{ATLAS2025,CMS:2022kqg}.

Starting from a final state with one top-quark fragmenting hadronically and one leptonically one would construct in a first step jets of hadrons for a given jet radius, say $R=0.2$ as in our $e^+e^-$-study above, in the hemisphere away to the hadronically decaying top-quark. Again, one of these jets is identified as a b-jet. Together with $e$ and $\nu$ one selects events with the mass $M(B e\nu)\sim m_t$  as candidates for events with gluon emission from top-quarks. Subsequently, a top-quark jet with opening $\Theta_{jet}$ is formed by combining the top-quark 
(the $B e\nu$-system) with the hadrons inside the cone $\Theta_{jet}$ out of the remaining jets of radius $R=0.2$ . The direction of the top-quark jet is found by maximizing the energy within the opening $\Theta_{jet}$. As in the case of the $e^+e^-$ process, the final $\xi$-distributions of hadrons for a given decay angle $X_b$ between the b and t-quark jets for the hemisphere opposite the b-quark jet are obtained and extrapolated to $X_b = 0$. A test of the extrapolation procedure can be obtained as in $e^+e^-$ annihilation by comparison with the stable top-quark at the parton level. Finally, the distribution $D_t(\xi,E_{jet};\Theta_{jet})$ for the hadrons is determined after a correction from the uncorrelated background has also been applied.

This result on the $\xi$-distribution can be compared with the corresponding result for the $\xi$-distribution in a light quark $uds$ jet at the same angle $\Theta_{jet}$ using the MCEG. The results should be similar to our Figs. \ref{fig:xi-dist-MLLA 1000}, \ref{fig:xi-dist-MLLA 1000-h1}, i.e. the ratio of distributions top/uds quarks should drop by about two orders of magnitude at small $\xi$ and so provide a very significant effect from the heavy quark mass.

Furthermore, it is possible to test the MLLA expectations for the $\xi$-distributions of particles in the top-quark jet as in Eq.~\eqref{DMLLA}, 
but the energy scale has to be redefined according to the reduced jet opening angle~\cite{Dokshitzer:1992jv,Dokshitzer1991Basics,Acosta:2003XYZ}
\begin{equation}
    W\to 2E_{jet} \sin \Theta_{jet}\;,
    \label{scale_with_angle}
\end{equation}
so we rewrite Eq.~\eqref{DMLLA} for the inclusive spectrum in a single jet with energy $E_{jet}$ at half opening angle $\Theta_{jet}$ 
\begin{equation}
    \bar D_Q(\xi,E_{jet};\Theta_{jet}) = \bar D_q(\xi,E_{jet}\sin(\Theta_{jet})) - \bar D_q(\xi - \xi_Q,\sqrt{e} m_Q).
    \label{DMLLA-1}
\end{equation} 
If the opening angle $\Theta_{jet}$ is reduced, the distribution of particles with high momenta or small $\xi$ that dominate at small angles remains approximately unchanged, but the production of low momentum particles at large $\xi$ is suppressed. 
Clearly, for $\Theta_{jet}\to \Theta_0$ the particle density vanishes in the approximation of Eq.~\eqref{DMLLA-1} and thus the relation can only be meaningful for sufficiently large jet angles or small $\Theta_0$, i.e.\ sufficiently high energies. The dependence of the characteristics of the $\xi$-spectra on the jet angle as in Eq.~\eqref{scale_with_angle} has been experimentally confirmed by the CDF collaboration \cite{Acosta:2003XYZ}.

The MLLA relation Eq.~\eqref{DMLLA-1} relates the $\xi$-distribution in top-quark production to the ones in light quark production at different energy scales. These distributions at the required higher energies in the TeV range have  not yet been measured. 
They have been determined at LEP energies (see i.e. our discussion for c- and b-quarks in \cite{Kluth:2023umf}) and can be extrapolated to higher energies relying on the QCD based MCEG. Alternatively, these distributions could also be obtained at the LHC in principle from the measurement of high $p_T$ direct photon production through subprocesses like $gq\to \gamma q$ involving light quarks $q$ with subsequent fragmentation of the light quarks  into hadrons.

As a second strategy to the dead cone studies we consider 
the application of substructure techniques of the top-quark jet
\cite{MaltoniSelvaggiThaler:2016DeadCone}. The method identifies the hard jet core of the top-quark by removing soft components of the jet.  The analysis starts from two “fat” jets where one contains a high $p_t$ lepton and a large missing transverse momentum. After removing the high $p_T$ lepton, the remaining particles in the fat jet are reclustered with the Cambridge-Aachen algorithm~\cite{Dokshitzer:1997in} to form an angular ordered tree. A “soft drop” algorithm~\cite{Larkoski:2014wba} isolates two subjets to become b- and g-jets. The final events should satisfy the mass constraint $M(b\ell\nu)\sim~m_t$ and also some cuts in the jet momenta $p_T^b> 50 $~GeV and $p_T^g>25$~GeV. By these cuts, part of the background from ISR and UE is removed as well, although not all of it as is concluded from a comparison with $e^+e^-$collisions. In this analysis the angular distribution  of gluons around the top-quark is compared with several MCEG's including and excluding the quark mass effect.

In comparing these two strategies we note: the first strategy attempts to correct for the background of low momentum particles to obtain the full $\xi$-fragmentation function of the top-quark, whereas the second strategy removes the low momentum particles from the sample. The latter method was applied with a focus on the  emission of gluons at small angles around the dead cone $\Theta\sim\Theta_0$. 
However, a cut in the gluon momenta would clearly also modify the $\xi$-spectra of hadrons at the larger $\xi$-values. In that case the behaviour of the $\xi$-spectrum at the smaller values would still show the large suppression in comparison to the light quark spectrum, which can be compared with the theory through the MCEG. A direct comparison with the MLLA prediction Eq.~\eqref{DMLLA-1} is possible within the first strategy, whereas within the second strategy the MLLA prediction had to be modified to consider low momentum cuts also to the light quark $\xi$-distributions for the two energy scales involved.

\section{Conclusions}

We have studied the feasibility of observing the dead cone in the forward emission of particles in a high energy top-quark jet, which is characteristic for the elementary QCD process of gluon emission from a massive quark. Our analysis is based on the application of the \textsc{Pythia} 8.3 MCEG for the events in the reaction $e^+e^-\to t\bar t +X$ at $\sqrt{s}=1$~TeV.

At first, we studied the angular distribution of particles around the top-quark direction that are bundled into the b-quark jet and the primary gluon jet as expected in the leading order of $\alpha_s$. In case of a stable top-quark one finds a strong suppression of gluon emission in the direction of the t-quark as expected in perturbative QCD. This ``dead cone" effect has already been observed in the production of  c- and b-quarks. A new phenomenon in top-quark production is the additional radiation generated by the b-quark from the top-quark decay, which partially obscures the dead cone phenomenon.  In leading order in $\alpha_s$ the additional radiation into the top-quark hemisphere comes from the $\widehat{tb}$ dipole which is superimposed on the $\widehat{t\bar t}$ dipole for stable top-quark production with dead cone and is characterized by the following properties:

(a) the gluons emitted in the direction of the b-quark are strongly limited in angle to 
half the top-quark hemisphere at the same side as the b-quark by ``angular ordering", as can be observed in the individual MC events and also in the 2-dim scatter plots;

(b) The gluons emitted from the $\widehat{tb}$ dipole in the t-quark direction prefer small angles and partially fill in the forward dead cone generated by the $\widehat{t\bar t} $ dipole; they are also radiated into the hemisphere opposite to the b-quark and their density increases with increasing b-quark decay angle; this is  against a naive expectation based on a picture of independent fragmentation of top- and b-quarks. The effect is enlarged for the parton and hadron final states which are studied in addition.

Our main aim is the study of momentum distributions. In view of the above results the momentum distribution of partons in the top-quark jet is then constructed from the partons in the half hemisphere opposite to the b-quark to remove contributions from $\widehat{tb}$ dipole radiation. To this end, the momentum distributions of partons are determined for different angles $\Theta_b$ of the b-quark jet and are extrapolated to angle $\Theta_b=0$ where the contribution of the $\widehat{tb}$ dipole radiation is expected to vanish. The resulting distribution for partons compares well with the corresponding distribution for the stable top-quark to within $5-15\%$ which we consider as a successful test of the method. The same procedure yields the momentum distribution of hadrons for the unstable top-quark and the related dead cone effect. 
This fragmentation function for the top-quark is found 
to be suppressed for large momenta or small $\xi=\ln{(1/x_p)}$,
at $\xi\sim 3$ by a factor down to $\sim1/100$, in comparison to the momentum distribution for the corresponding light quark jets. 

The  methods based on  the momentum and the angular studies both allow us to establish the dead cone effect. An advantage of the momentum study is the lower sensitivity to the top quark production angle and the possibility of subtracting the background from the $\widehat{tb}$ dipole by extrapolation in the b-quark decay angle. 
We also found that
the MLLA prediction, which relates the top-quark fragmentation function to the light quark fragmentation functions at c.m.s.\ energies $\sqrt{s}$ and $m_{t}$ is well satisfied by our analysis based on \textsc{Pythia} 8.3 within uncertainties around 10\%. This is in support of the construction of the jet final state within perturbative QCD, in particular the role of angular ordering within the evolution process at the large intrinsic scale of $m_t\sim 172$ GeV.  
Furthermore, we investigated the influence of the finite top-quark width on the particle spectra within a special option of \textsc{Pythia}. The effect exists for low momenta, but it is too small to be reliably resolved given our systematic uncertainty. 

Our analysis is performed for the forward hemisphere in the direction of the top-quark in $e^+e^-$ annihilation. This type of analysis also appears to be feasible for high $p_T$ top-quark production in pp collisions at the LHC, where the reduced jet cone angle of the top-quark jet as well as the influence from the underlying event, the initial state radiation and pile-up collisions have to be taken into account in the analysis.

\appendix

\section{Angular distributions of partons and hadrons in top-quark jets}
\label{app:angle}

\begin{figure}[thbp]
\begin{tabular}{cc}
\includegraphics[height=7.0cm,width=7.8cm]{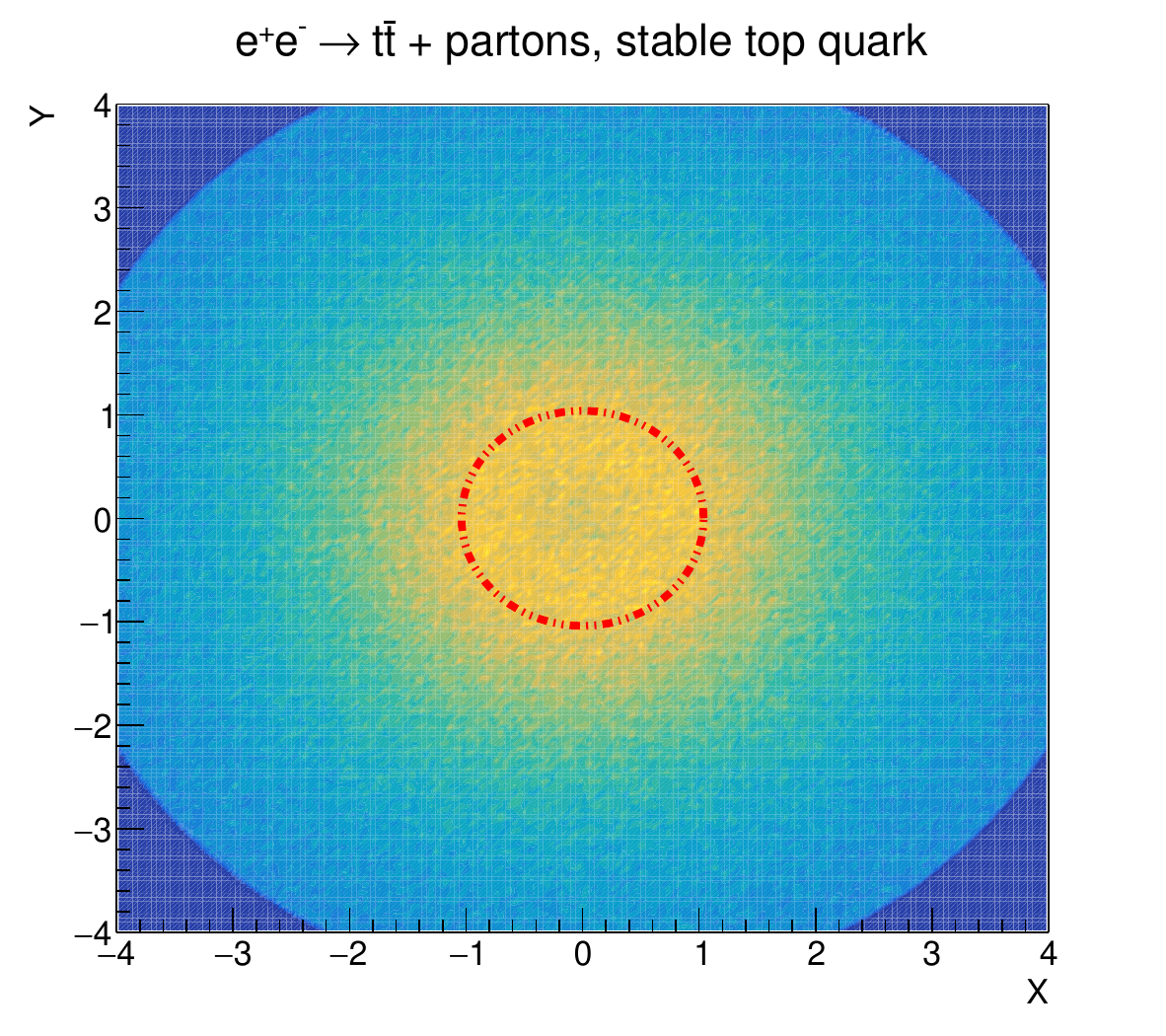} 
\hskip -0.5cm
\includegraphics[height=7.0cm,width=7.8cm]{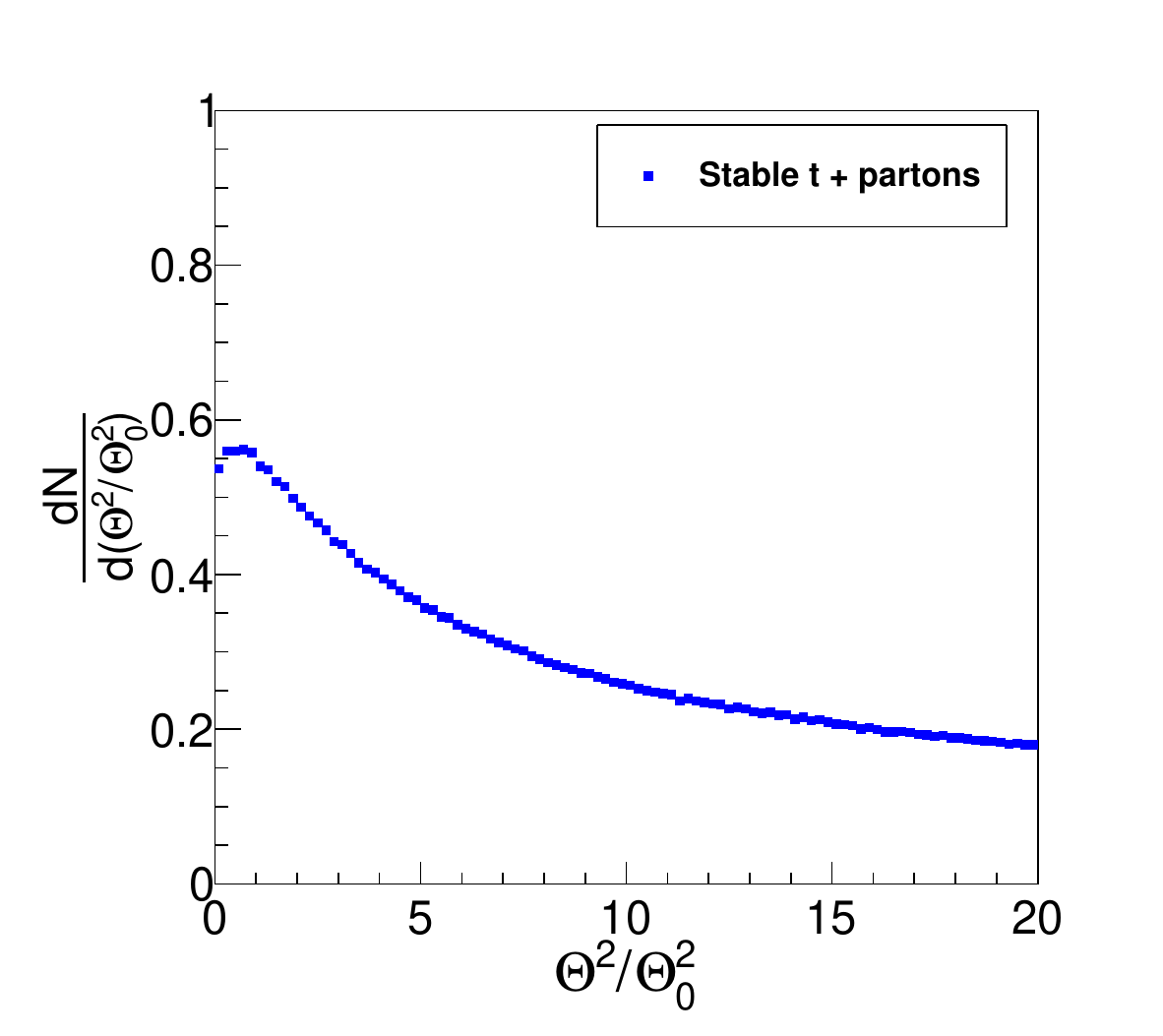}\\ 
\end{tabular}
  \caption{Angular distribution of partons around the direction of the stable top-quark showing a suppression of density in the center as in Fig.~\ref{fig:ring} but of smaller magnitude and distribution in the rescaled opening angles $\Theta/\Theta_0$ (top-quark excluded).}
  \label{fig:Theta2all}
\end{figure}

In this subsection we report on further studies of the angular structure of top-quark jets formed by partons and hadrons which evolve from the primary gluons. First we present in  Fig.~\ref{fig:Theta2all} the 2-dim ($X,Y$) distribution as in Fig.~\ref{fig:ring}, but for partons. The central dead cone is partially filled but still visible. Also shown is the distribution in the angle $\Theta^2/\Theta_0^2$ of the final partons themselves. This broadening of the distributions is expected from the convolution of two distributions. Therefore, the 
suppression near zero angle is weaker and the dead cone effect is reduced. 
\begin{figure}[b!]
\includegraphics[height=6cm,width=17cm]{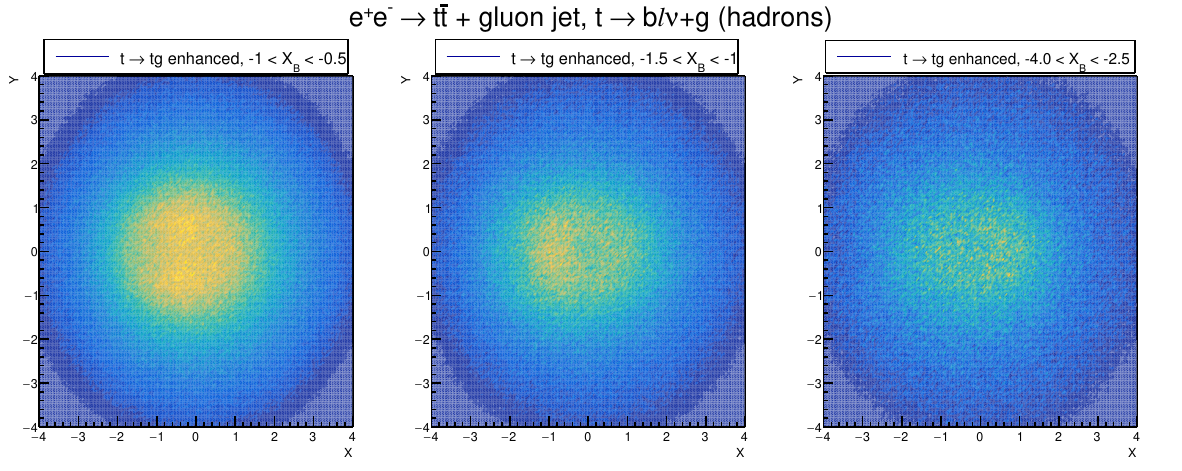} 
  \caption{Angular distribution of primary gluon jets as in Fig.~\ref{fig:scatter_gluon} but constructed from hadrons, for events enhanced with top-quark gluon emission $t\to t g$ for different angles $X_B$ of the b-quark jet to the top-quark direction.} 
  \label{fig:figure11up.pdf}
\end{figure}
\begin{figure}[htbp]
\includegraphics[height=6cm,width=17cm]{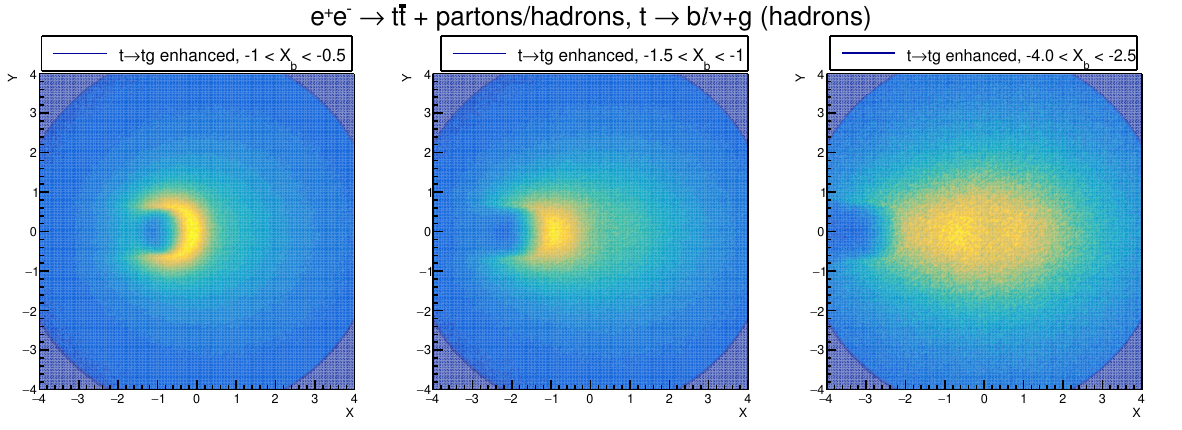} \\
\includegraphics[height=6cm,width=17cm]{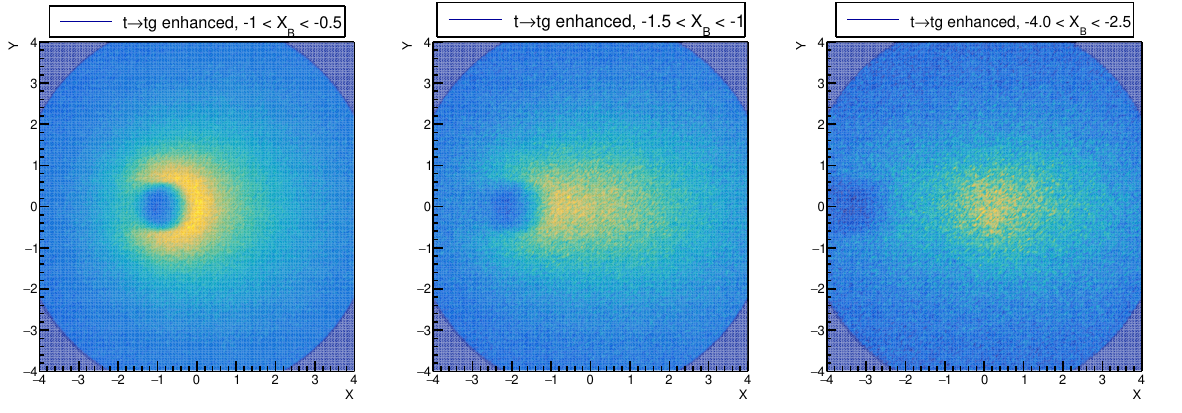} 
  \caption{Distributions over $(X,Y)$ angles of partons (upper row) and hadrons (lower row) for different angular regions of the b-quark jet $X_b$ or B-hadron jet $X_B$, all for events with top-quark gluon emission $t\to tg$ enhanced (vertical stripe in Fig.~\ref{fig:top_stripes}). }
  \label{fig:figure8-10up.pdf}
\end{figure}
\begin{figure} [htbp]
\centering
\includegraphics[width=0.52\linewidth]{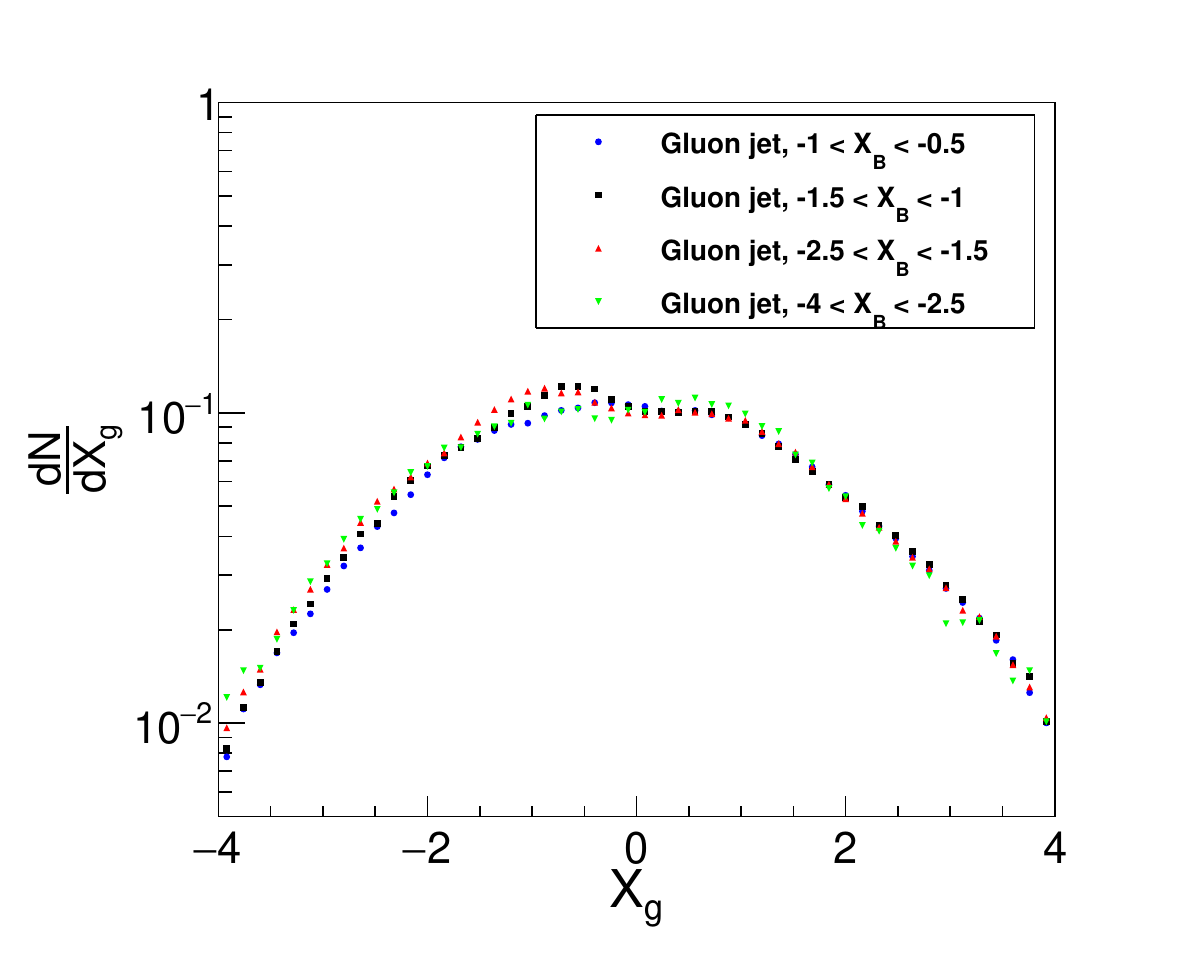}
\hskip -1.1cm
\includegraphics[width=0.52\linewidth]{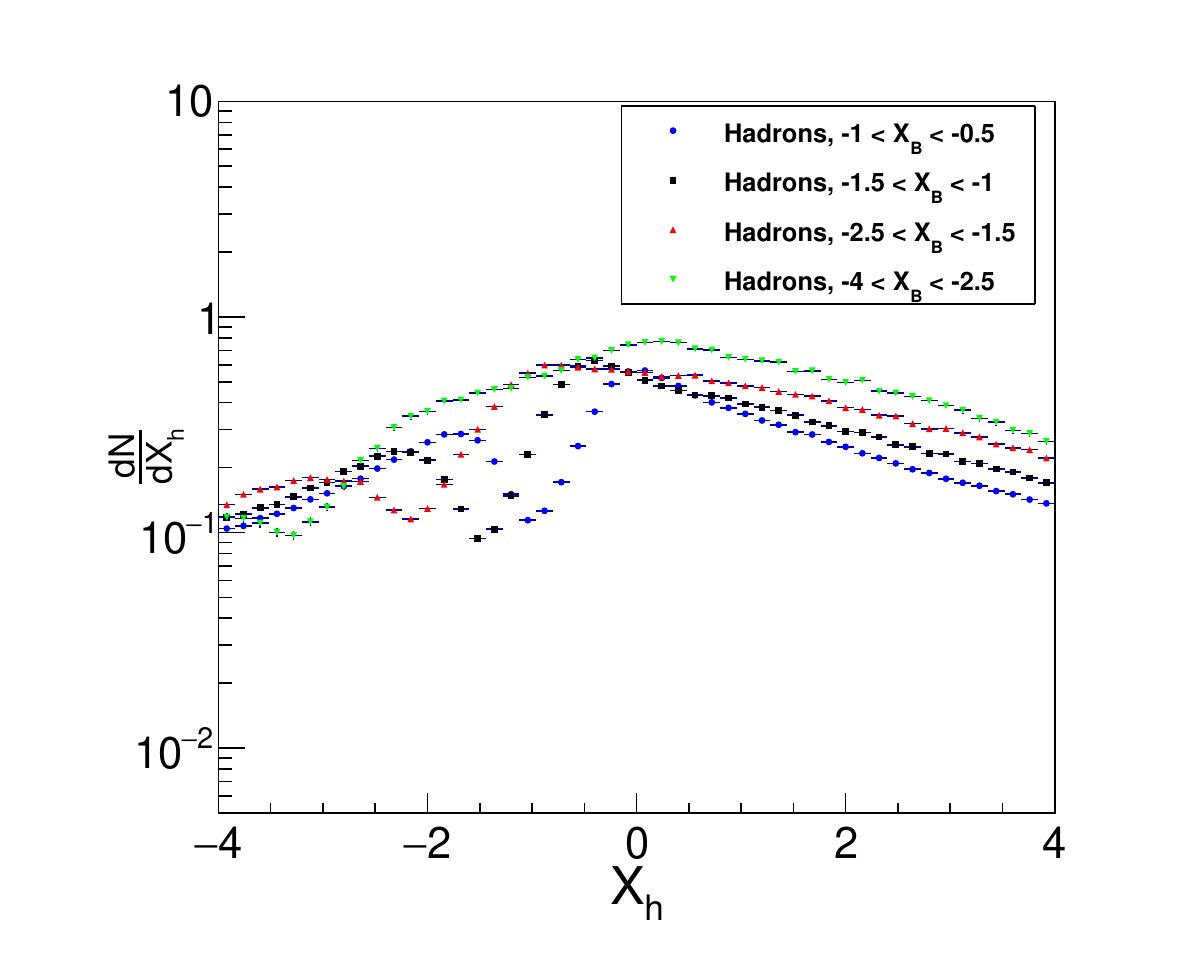}
  \caption{Distributions as in Fig.~\ref{fig:projections.gluon-parton} but for hadron final states: gluon jets over the angles $X_g$ and hadrons over $X_h$ along central stripe through $(X,Y)$ scatter plots in Fig.~\ref{fig:figure11up.pdf}, upper row and in Fig.~\ref{fig:figure8-10up.pdf}, lower row. The dependence of the spectra in the hemisphere $X_h>0$ on the B-jet decay angle $X_B$ is similar to the one for partons and reflects the $\widehat{tb}$ dipole contribution.}
  \label{fig:projections.parton}
\end{figure}

For the unstable top-quark we have constructed the angular distribution of primary gluons also from the hadronic final state in a similar way as we did for partons in Fig.~\ref{fig:scatter_gluon}, i.e.\ we assigned the hadrons to two jets, the one with a B-hadron is chosen as b-jet, the other one as primary gluon jet. 
The $(X,Y)$ angular distribution of these gluon jets constructed from hadrons is shown in Fig.~\ref{fig:figure11up.pdf}, where the events are chosen from the vertical stripe in Fig.~\ref{fig:top_stripes} (right panel) which enhances the process $t\to t g$. There is again some evidence of the dead cone surrounded by a circular radius distribution $R_X\sim 1$. The suppression of the b-jet signal inside the jet radius $R=0.2$ is only weakly visible here. This may indicate that the hadronization process in its MC realization yields a weaker collimation of gluons with their primary sources. On the other hand, for the angular distributions 
with t-quark decays enhanced, the half moon-type jet signature shows up  similarly to partons (not shown).

Finally we also present the angular distributions of the final state partons and hadrons for events with dominating $t\to tg$ process selected from the vertical stripes of Fig.~\ref{fig:top_stripes}. Their $(X,Y)$ angular distributions are shown in Fig.~\ref{fig:figure8-10up.pdf} using the same selection criteria for the masses. Again, the two different components of b- and t-quark radiation around the two centers are clearly visible;  however, the dead cone in the center is hardly visible for partons (upper row) and 
invisible for hadrons (lower row). We also note that at small angles $X_b$ for hadrons (lower left panel), the b-quark radiation extends slightly into the positive hemisphere $X>0$ in violation of the angular ordering valid at parton level. 

The distributions for primary gluon jets built from hadrons and for hadrons themselves along the central horizontal stripe in the $(X,Y)$ scatter plot, analogous to Fig.~\ref{fig:projections.gluon-parton}, are shown in Fig.~\ref{fig:projections.parton}. Here, the same ordering trends of the $X_g$ and $X_h$ distributions with angle $X_b$ 
are observed as for partons: the gluon jets are distributed rather independently of angle $X_b$  while the hadron density increases with increasing  $X_b$. 

\section{Distorted Gaussian with \(z^{5}\) and \(z^{6}\) Corrections}
\label{app:DG}

We define the dimensionless scaling variable
\begin{equation}
z \equiv \frac{\xi - {\xi_0}}{\sigma},
\end{equation}
where ${\xi_0}$ and $\sigma$ denote the position of the maximum and the width of the $\xi$-distribution, respectively.
The full distorted Gaussian distribution is given by
\begin{equation}
\label{eq:DGfull}
D(\xi)
=
\frac{N}{\sigma\sqrt{2\pi}}
\exp\!\left[
\mathrm{F}\!\left(
\frac{\xi-{\xi_0}}{\sigma}
\right)
\right],
\end{equation}
where the exponent $\mathrm{F(z)}$ is commonly expanded up to the 4th order
(see \cite{Fong:1990nt} for a review), for our fits terms up to the 6th order are required for a good fit
\begin{gather}
\label{eq:F_decomp}
\mathrm{F}(z) = -\frac{1}{2}z^{2}+\varepsilon(z)\\
\label{eq:eps_def}
\begin{aligned}
\varepsilon(z)={}&
\frac{k}{8}
-\frac{s}{2}z
-\frac{k}{4}z^{2}
+\frac{s}{6}z^{3}
+\frac{k}{24}z^{4}
+\frac{c_{5}}{120}z^{5}
+\frac{c_{6}}{720}z^{6}.
\end{aligned}
\end{gather}
Here we have separated the standard quadratic Gaussian part from the deformation $\epsilon$. $N$ is the overall normalization, corresponding to the particle multiplicity in the Gaussian limit;  $s$ controls the skewness and  $k$ he kurtosis of the distribution, $c_{5}$ and $c_{6}$ parameterize higher order shape deformations.
The even powers of $z$ determine the peak sharpness and tail structure, the odd powers generate the left right symmetry.
\begin{figure}[t]
    \centering
    \includegraphics[width=0.45\linewidth]{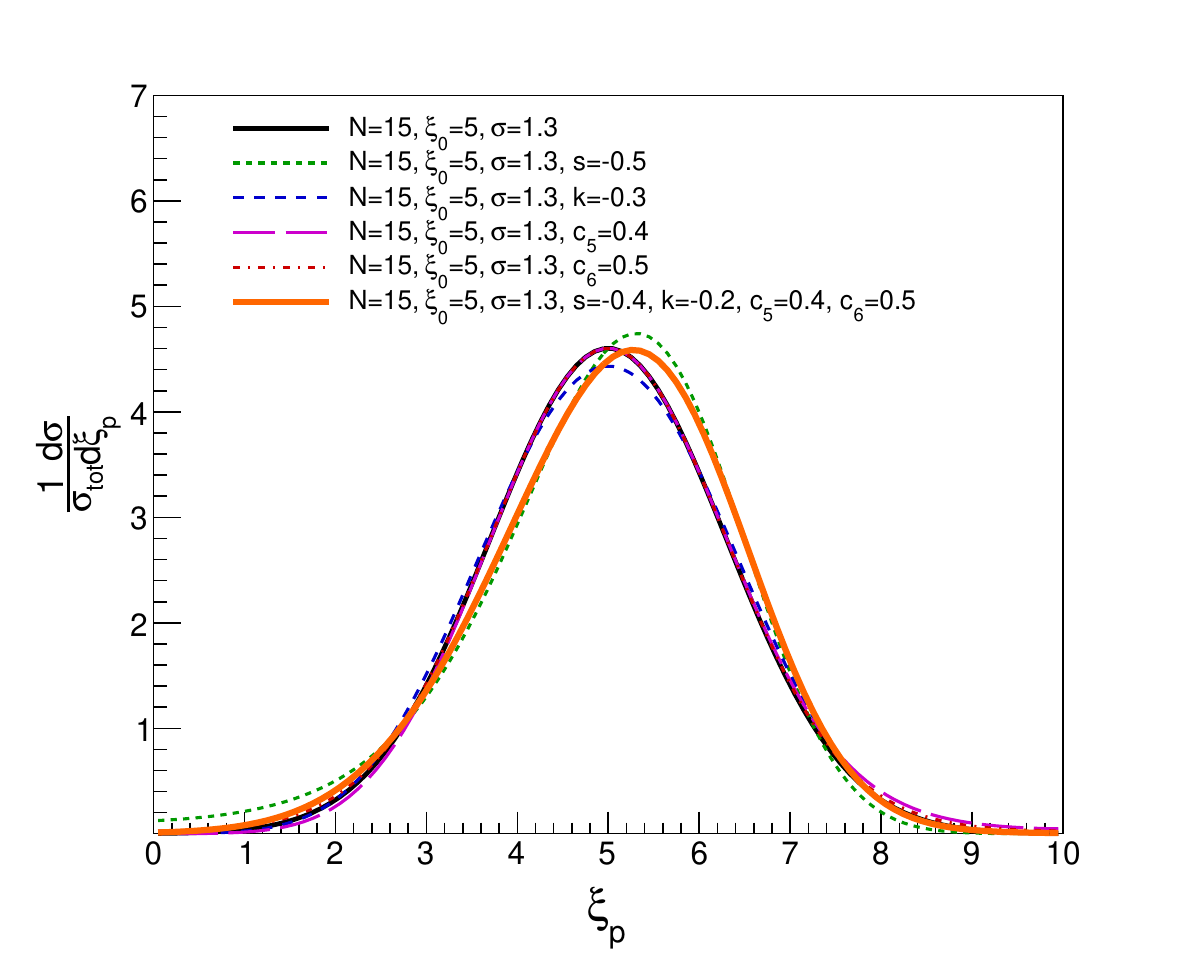}
    \includegraphics[width=0.45\linewidth]{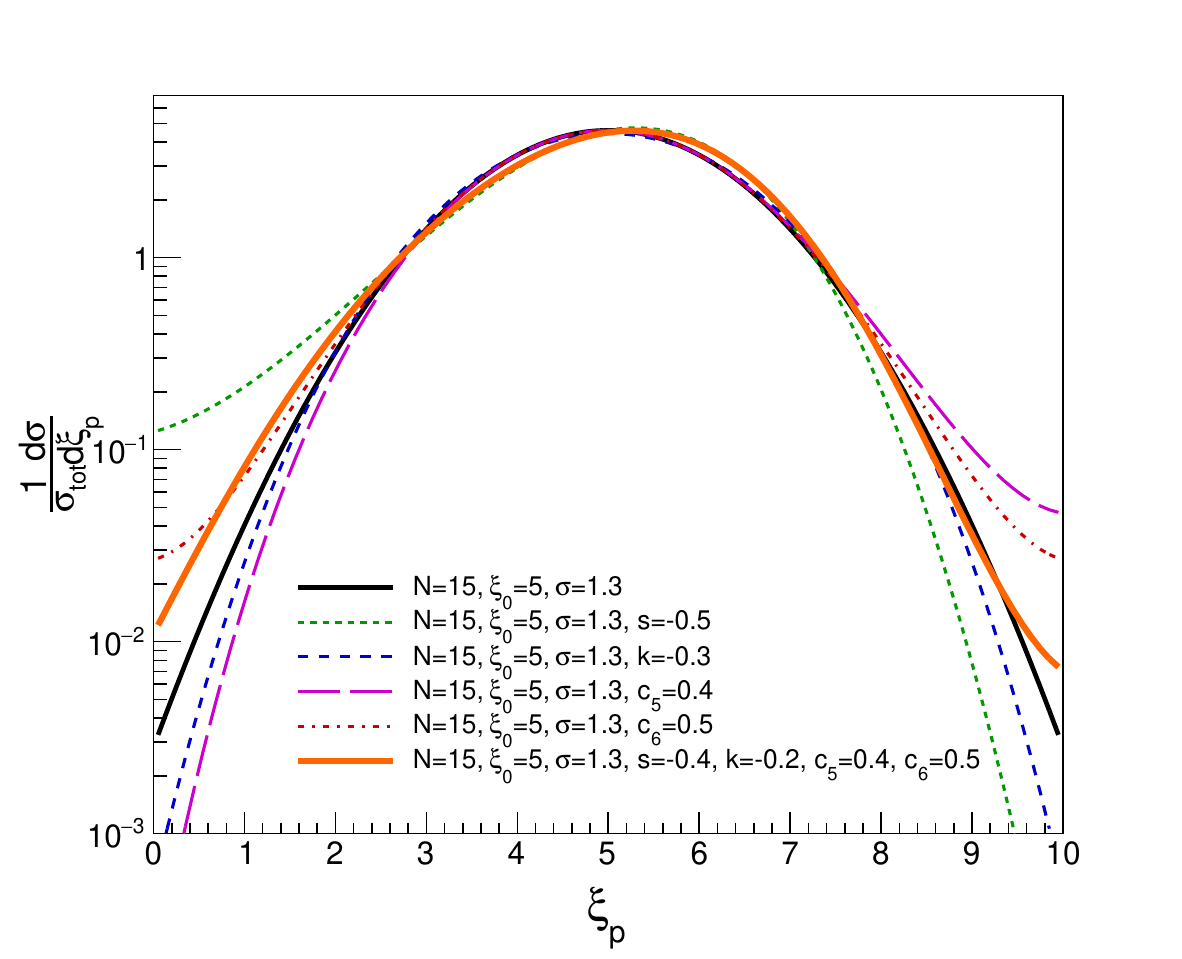}
\caption{
Illustration of the distorted Gaussian parametrization of the logarithmic
momentum distribution $D(\xi)$. The black curve shows the pure Gaussian limit
($s=k=c_{5}=c_{6}=0$). Each additional curve isolates the effect of a single
deformation parameter switched on individually: skewness ($s\neq0$, green),
kurtosis ($k\neq0$, blue), fifth order deformation ($c_{5}\neq0$, magenta),
and sixth order tail correction ($c_{6}\neq0$, red).
The numerical values are chosen for illustration close to our fit results.
}
\label{fig:DGmoments}
\end{figure}
An illustration of the role of the various parameters in the shape of the $\xi$-distribution is presented in Fig.~\ref{fig:DGmoments}. 

\subsection*{Small-deformation expansion}

For small deviations from Gaussianity we expand $e^{\varepsilon(z)} \simeq 1+\varepsilon(z)$, 
so that
\begin{equation}
\label{eq:norm_linear_eps}
\int_{-\infty}^{+\infty} D(\xi)\, d\xi
\simeq
N\left(1+\langle \varepsilon(z)\rangle_G\right).
\end{equation}
where we introduced the Gaussian-weighted average
\begin{equation}
\label{eq:gauss_avg}
\langle g(z)\rangle_{G}
\equiv
\frac{1}{\sqrt{2\pi}}
\int_{-\infty}^{+\infty} g(z)\,e^{-z^{2}/2}\,dz.
\end{equation}
Using standard Gaussian moments with respect to $\langle\cdot\rangle_G$,
\[
\langle z\rangle_G=0,\quad
\langle z^{2}\rangle_G=1,\quad
\langle z^{3}\rangle_G=0,\quad
\langle z^{4}\rangle_G=3,\quad
\langle z^{5}\rangle_G=0,\quad
\langle z^{6}\rangle_G=15,
\]
one finds, to linear order in the deformation parameters,
\begin{equation}
\label{eq:eps_avg_result}
\langle \varepsilon(z)\rangle_G
=
\frac{k}{8}-\frac{k}{4}+\frac{k}{8}
+\frac{15}{720}c_{6}
=
\frac{c_{6}}{48}.
\end{equation}
Therefore, to leading order,
\begin{equation}
\label{eq:norm_final}
\int_{-\infty}^{+\infty} D(\xi)\, d\xi
=
N\left(
1+\frac{c_{6}}{48}
+\mathcal{O}(s^{2},k^{2},c_{5}^{2},c_{6}^{2})
\right).
\end{equation}
In particular, if $s=k=c_{5}=c_{6}=0$, the DG distribution reduces to a Gaussian
and the integral equals $N$ exactly.

\section{Kinematics}
\label{app:kinematics}

We define the three momenta for the top-quark, parton, and $e^-$ beam as
\begin{equation}
    \mathbf{p}^{\,t}=(p_x^{t},\,p_y^{t},\,p_z^{t}),\qquad
    \mathbf{p}=(p_x,\,p_y,\,p_z),\qquad
    \mathbf{p}^{\,e^-}=(p_x^{e^-},\,p_y^{e^-},\,p_z^{e^-}).
\end{equation}
The vector normals to the plane spanned by the top-quark and the parton or the hadron, respectively, are
\begin{equation}
\mathbf{p}^{\,t,p}_{\perp}
=\mathbf{p}^{\,t}\times\mathbf{p}, \qquad 
 \mathbf{p}^{\,t,e^-}_{\perp}
    =\mathbf{p}^{\,t}\times\mathbf{p}^{\,e^-}.
\end{equation}
The azimuthal angle $\phi$ between the two planes is defined through
\begin{equation}
\cos\phi
=
\frac{
\mathbf{p}^{\,t,p}_{\perp}\cdot\mathbf{p}^{\,t,e^-}_{\perp}
}{
\lvert\mathbf{p}^{\,t,p}_{\perp}\rvert\,
\lvert\mathbf{p}^{\,t,e^-}_{\perp}\rvert
}.
\end{equation}
and its orientation is fixed by the sign of
\begin{equation}
s=(\mathbf{p}^{\,t,p}_{\perp}\times\mathbf{p}^{\,t,e^-}_{\perp})
\cdot
\mathbf{p}^{\,t},
\end{equation}
such that
\[
\phi=
\begin{cases}
\arccos(\cos\phi), & s>0,\\[6pt]
2\pi-\arccos(\cos\phi), & s<0.
\end{cases}
\]
The polar angle $\Theta$ between the parton and the top-quark momentum
is given by
\begin{equation}
\Theta
=
\arccos\!\left(
\frac{\mathbf{p}^{\,t}\cdot\mathbf{p}}
{\lvert\mathbf{p}^{\,t}\rvert\,\lvert\mathbf{p}\rvert}
\right).
\end{equation}
Finally, this yields the coordinates $X$ and $Y$ in Eq. \eqref{XYdef} in the plane perpendicular to the top-quark direction used for our angular scatter plots. In these plots, we present the distributions after rotation around the top-quark direction, so that the B hadron points into the negative X direction. This is achieved by replacing all azimuthal angles $\phi$ by  
\begin{equation}
\phi' = \phi - \phi_B - \pi  .
\end{equation}
\bibliography{hbpdeadcone}

\end{document}